\documentclass[twocolumn,10pt,letterpaper,superscriptaddress,amsmath,amssymb,
aip,
pop,
floatfix,
showkeys]{revtex4-2}

\usepackage{graphicx}
\usepackage{booktabs}
\usepackage{dcolumn}
\usepackage{bm}
\usepackage{float}
\usepackage[T1]{fontenc}
\usepackage{xcolor}
\usepackage{textcomp,gensymb} 
\usepackage{lineno}
\usepackage{soul}
\usepackage{ragged2e}
\usepackage{hyperref} 
\hypersetup{colorlinks=true,linkcolor=blue,filecolor=magenta,urlcolor=cyan}
\newcounter{para}

\begin{document}

\title{Characterization and performance of the Apollon main short-pulse laser beam following its commissioning at 2 PW level}

\author{Weipeng Yao} 
\email{weipeng.yao@obspm.fr}
\affiliation{LULI - CNRS, CEA, Sorbonne Universit\'e, Ecole Polytechnique, Institut Polytechnique de Paris - F-91128 Palaiseau cedex, France}
\affiliation{Sorbonne Universit\'e, Observatoire de Paris, Universit\'e PSL, CNRS, LERMA, F-75005, Paris, France}

\author{Ronan Lelièvre} 
\affiliation{LULI - CNRS, CEA, Sorbonne Universit\'e, Ecole Polytechnique, Institut Polytechnique de Paris - F-91128 Palaiseau cedex, France}
\affiliation{Laboratoire de micro-irradiation, de métrologie et de dosimétrie des neutrons, PSE-Santé/SDOS, IRSN, 13115 Saint-Paul-Lez-Durance, France}

\author{Itamar Cohen} 
\affiliation{The School of Physics and Astronomy, Tel Aviv University, Tel-Aviv, 6997801, Israel}
\affiliation{Light Stream Labs LLC, USA, Palo Alto, CA 94306}

\author{Tessa Waltenspiel} 
\affiliation{LULI - CNRS, CEA, Sorbonne Universit\'e, Ecole Polytechnique, Institut Polytechnique de Paris - F-91128 Palaiseau cedex, France}
\affiliation{University of Bordeaux, Centre Lasers Intenses et Applications, CNRS, CEA, UMR 5107, F-33405 Talence, France}
\affiliation{INRS-EMT, 1650 boul, Lionel-Boulet, Varennes, QC, J3X 1S2, Canada}

\author{Amokrane Allaoua} 
\affiliation{Laboratoire de micro-irradiation, de métrologie et de dosimétrie des neutrons, PSE-Santé/SDOS, IRSN, 13115 Saint-Paul-Lez-Durance, France}

\author{Patrizio Antici} 
\affiliation{INRS-EMT, 1650 boul, Lionel-Boulet, Varennes, QC, J3X 1S2, Canada}

\author{Yohan Ayoul} 
\affiliation{LULI - CNRS, CEA, Sorbonne Universit\'e, Ecole Polytechnique, Institut Polytechnique de Paris - F-91128 Palaiseau cedex, France}

\author{Arie Beck}
\affiliation{Physics Department, Nuclear Research Center - Negev, Beer-Sheva, 84190, Israel}

\author{Audrey Beluze} 
\affiliation{LULI - CNRS, CEA, Sorbonne Universit\'e, Ecole Polytechnique, Institut Polytechnique de Paris - F-91128 Palaiseau cedex, France}

\author{Christophe Blancard} 
\affiliation{CEA, DAM, DIF, F-91297 Arpajon, France}
\affiliation{Universit\'{e} Paris-Saclay, CEA, LMCE, 91680 Bruy\`{e}res-le-Ch\^{a}tel, France}

\author{Daniel Cavanna} 
\affiliation{LULI - CNRS, CEA, Sorbonne Universit\'e, Ecole Polytechnique, Institut Polytechnique de Paris - F-91128 Palaiseau cedex, France}

\author{Mélanie Chabanis} 
\affiliation{LULI - CNRS, CEA, Sorbonne Universit\'e, Ecole Polytechnique, Institut Polytechnique de Paris - F-91128 Palaiseau cedex, France}

\author{Sophia N. Chen} 
\affiliation{``Horia Hulubei'' National Institute for Physics and Nuclear Engineering, 30 Reactorului Street, RO-077125, Bucharest-Magurele, Romania}

\author{Erez Cohen}
\affiliation{Physics Department, Nuclear Research Center - Negev, Beer-Sheva, 84190, Israel}

\author{Quentin Ducasse} 
\affiliation{Laboratoire de micro-irradiation, de métrologie et de dosimétrie des neutrons, PSE-Santé/SDOS, IRSN, 13115 Saint-Paul-Lez-Durance, France}

\author{Mathieu Dumergue} 
\affiliation{LULI - CNRS, CEA, Sorbonne Universit\'e, Ecole Polytechnique, Institut Polytechnique de Paris - F-91128 Palaiseau cedex, France}

\author{Fouad El Hai} 
\affiliation{LULI - CNRS, CEA, Sorbonne Universit\'e, Ecole Polytechnique, Institut Polytechnique de Paris - F-91128 Palaiseau cedex, France}

\author{Christophe Evrard} 
\affiliation{LULI - CNRS, CEA, Sorbonne Universit\'e, Ecole Polytechnique, Institut Polytechnique de Paris - F-91128 Palaiseau cedex, France}

\author{Evgeny Filippov} 
\affiliation{CLPU, 37185, Villamayor, Salamanca, Spain}

\author{Antoine Freneaux} 
\affiliation{LULI - CNRS, CEA, Sorbonne Universit\'e, Ecole Polytechnique, Institut Polytechnique de Paris - F-91128 Palaiseau cedex, France}

\author{Donald Cort Gautier} 
\affiliation{LANL, PO Box 1663, Los Alamos, New Mexico 87545, USA}

\author{Fabrice Gobert} 
\affiliation{LULI - CNRS, CEA, Sorbonne Universit\'e, Ecole Polytechnique, Institut Polytechnique de Paris - F-91128 Palaiseau cedex, France}

\author{Franck Goupille} 
\affiliation{LULI - CNRS, CEA, Sorbonne Universit\'e, Ecole Polytechnique, Institut Polytechnique de Paris - F-91128 Palaiseau cedex, France}

\author{Michael Grech} 
\affiliation{LULI - CNRS, CEA, Sorbonne Universit\'e, Ecole Polytechnique, Institut Polytechnique de Paris - F-91128 Palaiseau cedex, France}

\author{Laurent Gremillet} 
\affiliation{CEA, DAM, DIF, F-91297 Arpajon, France}
\affiliation{Universit\'{e} Paris-Saclay, CEA, LMCE, 91680 Bruy\`{e}res-le-Ch\^{a}tel, France}

\author{Yoav Heller} 
\affiliation{LULI - CNRS, CEA, Sorbonne Universit\'e, Ecole Polytechnique, Institut Polytechnique de Paris - F-91128 Palaiseau cedex, France}

\author{Emmanuel d'Humi\`eres} 
\affiliation{University of Bordeaux, Centre Lasers Intenses et Applications, CNRS, CEA, UMR 5107, F-33405 Talence, France}

\author{Hanna Lahmar} 
\affiliation{LULI - CNRS, CEA, Sorbonne Universit\'e, Ecole Polytechnique, Institut Polytechnique de Paris - F-91128 Palaiseau cedex, France}

\author{Livia Lancia} 
\affiliation{LULI - CNRS, CEA, Sorbonne Universit\'e, Ecole Polytechnique, Institut Polytechnique de Paris - F-91128 Palaiseau cedex, France}

\author{Nathalie Lebas} 
\affiliation{LULI - CNRS, CEA, Sorbonne Universit\'e, Ecole Polytechnique, Institut Polytechnique de Paris - F-91128 Palaiseau cedex, France}

\author{Ludovic Lecherbourg } 
\affiliation{CEA, DAM, DIF, F-91297 Arpajon, France}
\affiliation{Universit\'{e} Paris-Saclay, CEA, LMCE, 91680 Bruy\`{e}res-le-Ch\^{a}tel, France}

\author{Stéphane Marchand} 
\affiliation{LULI - CNRS, CEA, Sorbonne Universit\'e, Ecole Polytechnique, Institut Polytechnique de Paris - F-91128 Palaiseau cedex, France}

\author{Damien Mataja} 
\affiliation{LULI - CNRS, CEA, Sorbonne Universit\'e, Ecole Polytechnique, Institut Polytechnique de Paris - F-91128 Palaiseau cedex, France}

\author{Gabriel Meyniel} 
\affiliation{LULI - CNRS, CEA, Sorbonne Universit\'e, Ecole Polytechnique, Institut Polytechnique de Paris - F-91128 Palaiseau cedex, France}

\author{David Michaeli}
\affiliation{Physics Department, Nuclear Research Center - Negev, Beer-Sheva, 84190, Israel}

\author{Dimitris Papadopoulos} 
\affiliation{LULI - CNRS, CEA, Sorbonne Universit\'e, Ecole Polytechnique, Institut Polytechnique de Paris - F-91128 Palaiseau cedex, France}

\author{Frédéric Perez} 
\affiliation{LULI - CNRS, CEA, Sorbonne Universit\'e, Ecole Polytechnique, Institut Polytechnique de Paris - F-91128 Palaiseau cedex, France}

\author{Sergy Pikuz} 
\affiliation{HB11 Energy Holdings, Freshwater, NSW 2096, Australia} 

\author{Ishay Pomerantz} 
\affiliation{The School of Physics and Astronomy, Tel Aviv University, Tel-Aviv, 6997801, Israel}

\author{Patrick Renaudin} 
\affiliation{CEA, DAM, DIF, F-91297 Arpajon, France}
\affiliation{Universit\'{e} Paris-Saclay, CEA, LMCE, 91680 Bruy\`{e}res-le-Ch\^{a}tel, France} 

\author{Lorenzo Romagnani} 
\affiliation{LULI - CNRS, CEA, Sorbonne Universit\'e, Ecole Polytechnique, Institut Polytechnique de Paris - F-91128 Palaiseau cedex, France}

\author{François Trompier}
\affiliation{Laboratoire de dosim\'etrie des rayonnements ionisants, PSE-Sant\'e/SDOS, IRSN, 92262 Fontenay-aux-Roses, France}

\author{Edouard Veuillot} 
\affiliation{LULI - CNRS, CEA, Sorbonne Universit\'e, Ecole Polytechnique, Institut Polytechnique de Paris - F-91128 Palaiseau cedex, France}

\author{Thibaut Vinchon} 
\affiliation{Laboratoire de micro-irradiation, de métrologie et de dosimétrie des neutrons, PSE-Santé/SDOS, IRSN, 13115 Saint-Paul-Lez-Durance, France}

\author{François Mathieu} 
\affiliation{LULI - CNRS, CEA, Sorbonne Universit\'e, Ecole Polytechnique, Institut Polytechnique de Paris - F-91128 Palaiseau cedex, France}

\author{Julien Fuchs} 
\email{julien.fuchs@polytechnique.fr}
\affiliation{LULI - CNRS, CEA, Sorbonne Universit\'e, Ecole Polytechnique, Institut Polytechnique de Paris - F-91128 Palaiseau cedex, France}

\date{\today}

\begin{abstract}

We present the results of the second commissioning phase of the short-focal-length area of the Apollon laser facility (located in Saclay, France), which was performed with the main laser beam (F1), scaled to a peak power of 2 PetaWatt. Under the conditions that were tested, this beam delivered on-target pulses of maximum energy up to 45 J and 22 fs duration. Several diagnostics were fielded to assess the performance of the facility. The on-target focal spot and its spatial stability, as well as the secondary sources produced when irradiating solid targets, have all been characterized, with the goal of helping users design future experiments. The laser-target interaction was characterized, as well as emissions of energetic ions, X-ray and neutrons recorded, all showing good laser-to-target coupling efficiency. Moreover, we demonstrated the simultaneous fielding of F1 with the auxiliary 0.5 PW F2 beam of Apollon, enabling dual beam operation. The present commissioning will be followed in 2025 by a further commissioning stage of F1 at the 8 PW level, en route to the final 10 PW goal.

\end{abstract}


\maketitle

\bigskip

\section{Introduction}
\label{INTRO}

With the development of laser technology, high-power lasers have become indispensable tools to investigate extreme states of matter subject to ultra-strong electromagnetic fields, enabling a plethora of scientific and technical applications \cite{daido2012review, danson2019petawatt, albert20212020}, including the generation of unprecedentedly dense beams of energetic particles, the development of ultra-short and/or ultra-bright photon and neutron sources, and the laboratory reproduction of high-energy astrophysical events. 

\begin{figure*}[ht]
    \centering
    \includegraphics{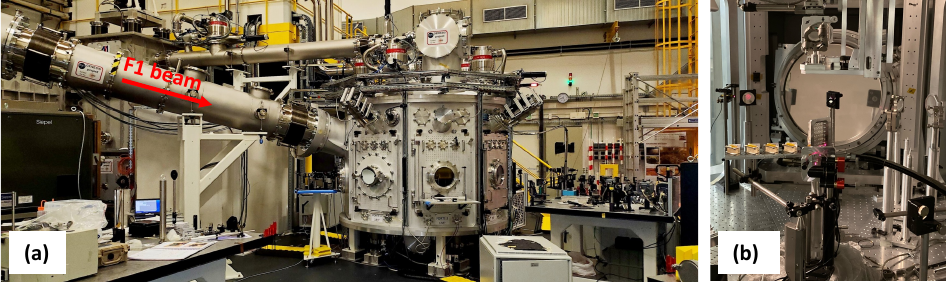}
    \caption{Photographs of the (a) Apollon SFA with the F1 beamline; (b) the 500 mm diameter off-axis parabolic (OAP) of the F1 beam inside the vacuum chamber.}
    \label{fig:photo}
\end{figure*}

\begin{figure*}[ht]
    \centering
    \includegraphics[width=0.9\textwidth]{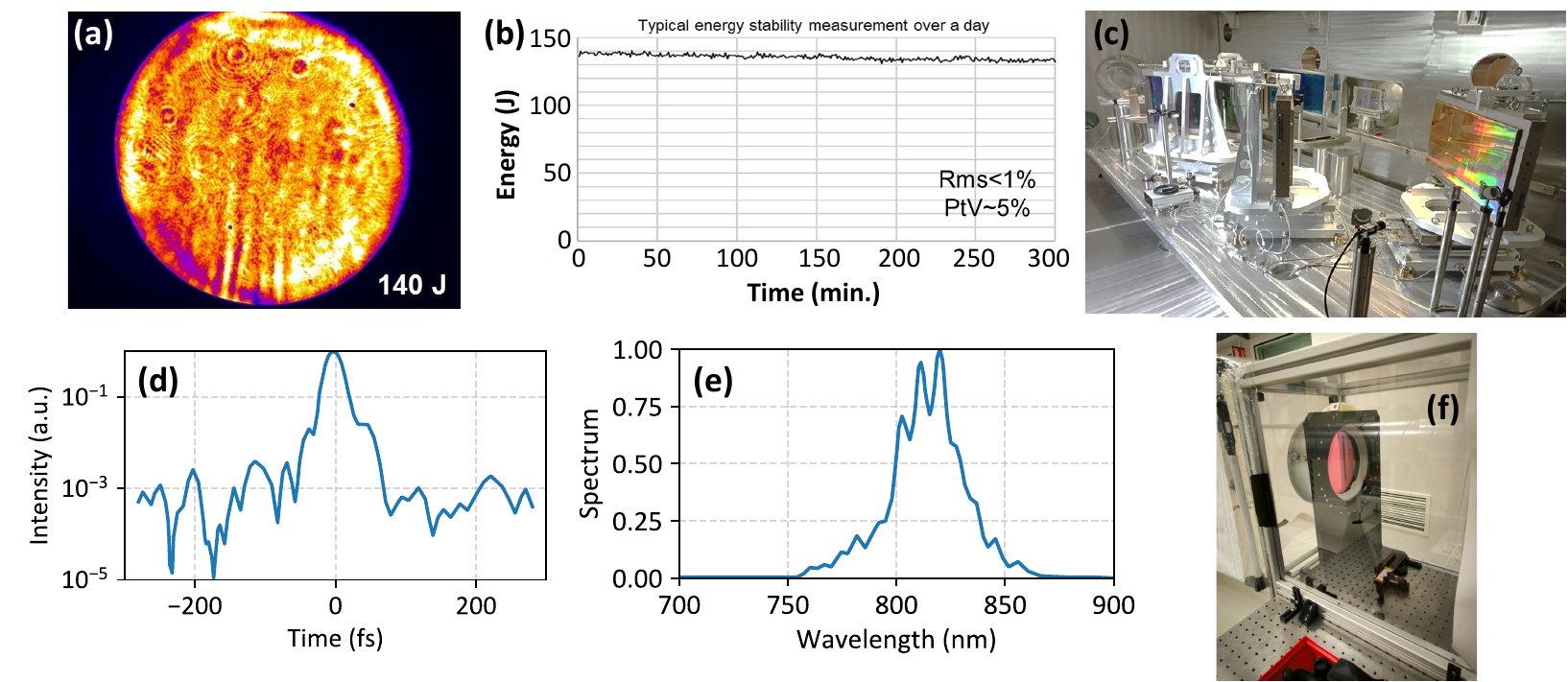}
    \caption{(a) Near-field energy distribution of the F1 beam. (b) Typical energy stability of the last amplifier over 300 shots (5 hours).  (c) General view of the inside of the 10 PW compressor of the F1 beamline (the size of each grating is $910\times 455$ mm$^2$). 
    (d) Wizzler measurement of the compressed pulses at the output of the 10 PW compressor, showing a pulse duration of 23.7 fs (for a 23.1 fs Fourier transform limited duration). (e) The same Wizzler measurement in wavelength. (f) Photo of Apollon's last amplifier.}
    \label{fig:laser}
\end{figure*}

The Apollon laser facility (Orme des Merisiers, Saclay, France) aims to be among the few multi-PW installations in the world devoted to the study of high intensity laser matter interaction in unprecedented regimes and at peak intensities in the $>10^{23}\,\rm W\,cm^{-2}$ range. The final goal of the Apollon laser is the generation of 10 PW peak power pulses corresponding to an energy of 150 J and 15 fs duration at a repetition rate of 1 shot/minute \cite{papadopoulos2016apollon,zou2015design}. Apollon provides up to four beam lines (F1: 10 PW, F2: 1 PW, F3: the uncompressed beam and F4: 10 TW probe beam), all generated by the same beam after the last amplifier, with the possibility to be combined on target under different geometries and synchronization configurations. The secondary beam F2, with a maximum power of 1 PW, has been commissioned in 2021 \cite{burdonov2021characterization} and simultaneously available to users. 
The Apollon facility offers two experimental areas available to the users: (1) The Short Focus Area (SFA), where tight focusing (F\#2.5) on solid targets is the principal objective; (2) The Long Focus Area (LFA), where mostly gas targets and wakefield electron acceleration experiments are realized. The Apollon facility is open to the international research community since 2021 with about 15 experimental campaigns already realized with the 1 PW beamline of the system \cite{papadopoulos2019first,ranc2020improvement,burdonov2021characterization,bleotu2022spectral,moulanier2023modeling,lelievre2023comprehensive,lahaye2024laser,yao2024enhanced}.

In this paper, we report on the current status of the Apollon F1 beam and present the results of its commissioning in the SFA in April 2023 with a 2 PW peak power level, and the first experimental demonstration of its combined use with F2. This first experiment with F1 was devoted to characterize the laser-target coupling, the subsequent laser-driven ion acceleration from a solid target, and neutron generation from a secondary converter. 
Fig.~\ref{fig:photo} (a) shows an overview of the SFA with the F1 beamline. Fig.~\ref{fig:photo} (b) displays the parabolic mirror of the F1 beam inside the target chamber, together with some of the equipments fielded during the commissioning campaign. 

The paper is organized as follows: in Sec.~\ref{CHA}, we present the characterized F1 beam parameters; in Sec.~\ref{sec:setup}, we show the experimental setup and the characterization of the plasma conditions; in Sec.~\ref{protons}, we present the characterization of the hot-electron generation and ion acceleration; in Sec.~\ref{neutron}, the characterization of the neutron generation will be presented; in Sec.~\ref{x-ray_radio}, we report on the capability of X-ray generation and radiography; and in Sec.~\ref{F2}, we discuss the simultaneous operation of F1 with F2. In Sec.~\ref{prospects}, we discuss the perspectives and at last we conclude in Sec.~\ref{conclusion}.

\section{Characteristics of the main beamline of Apollon}
\label{CHA}

After the activation of Apollon's fifth and final amplification stage, the F1 beam has been commissioned to an intermediate pulse energy of 140 J. Six separate pump beamlines are used to deliver a total of 270 Joules on the 196 mm diameter Ti:Sa crystal of the amplifier, shown in Fig.~\ref{fig:laser} (f). A typical fluence distribution of the near-field of the amplified beam is shown in Fig.~\ref{fig:laser} (a), presenting a flat-top repartition and a good homogeneity over the full aperture of 140 mm in diameter. The amplifier has been extensively tested regarding its short and mid-term stability characteristics showing typically <5\% PtV energy fluctuation over a complete day of operation, i.e., 300 shots, over 5 hours, as shown in Fig.~\ref{fig:laser} (b). The pointing and wavefront stability has also been measured with about 7 $\mu$rad PtV and 70$\pm$10\% Strehl ratio fluctuations respectively over 30 full energy shots. The beam is then expanded in an off-axis parabolic (OAP) telescope to 400 mm diameter and sent for compression in the 10 PW beamline compressor of Apollon, shown in Fig.~\ref{fig:laser} (c). The compressed pulses, with a spectrum centered at $\lambda = 815$ nm and extending over the 755-875 nm range, are measured to a <24 fs pulse duration based on a Wizzler device, as shown in Fig.~\ref{fig:laser} (d) and (e). For this first commissioning, the temporal contrast of the pulses could not be fully characterized, but only measured with the low energy beam (at 10 Hz) showing  the same characteristics as for the F2 beamline \cite{burdonov2021characterization,ranc2020improvement}. Careful further investigation is required to precisely evaluate the level of pre-pulses both in the picosecond range (under the influence of the full non-linearity of the amplified beam) as well as in the nanosecond long range for potential diffusion contributions from the crystal of the last amplification stage.

Based on the transmission of the compressor, measured to be $\sim$70\%, together with the beam transport between the amplifier’s output and the compressor ($\sim$92\%), we estimate the maximum energy of the pulses, at the exit of the compressor, to be >90 Joules, corresponding therefore to >3.7 PW peak power capacity. The compressed beam has been then sent to the SFA area, where focusing is based on an OAP of 1 m focal distance (F\#2.5), as shown in Fig.~\ref{fig:photo} (b).

The measured total laser chain transmission (between the amplifier and the target) is $\sim$55\%. For the SFA commissioning experiment presented in this work we used a non-optimized and uncoated protection ultra-thin film ($\sim 3\  \mu \rm m$) installed directly in front of the focusing parabola in double pass configuration for the focused beam. The uncoated surfaces, but mostly the diffusion of the protection film, resulted in about 40\% of additional losses and limited the actual energy on the target to <45 Joules. Therefore, the effective on-target peak power was around 2 PW. The quality of the focal spot however was intact.

\begin{figure}[ht]
    \centering
    \includegraphics[width=0.42\textwidth]{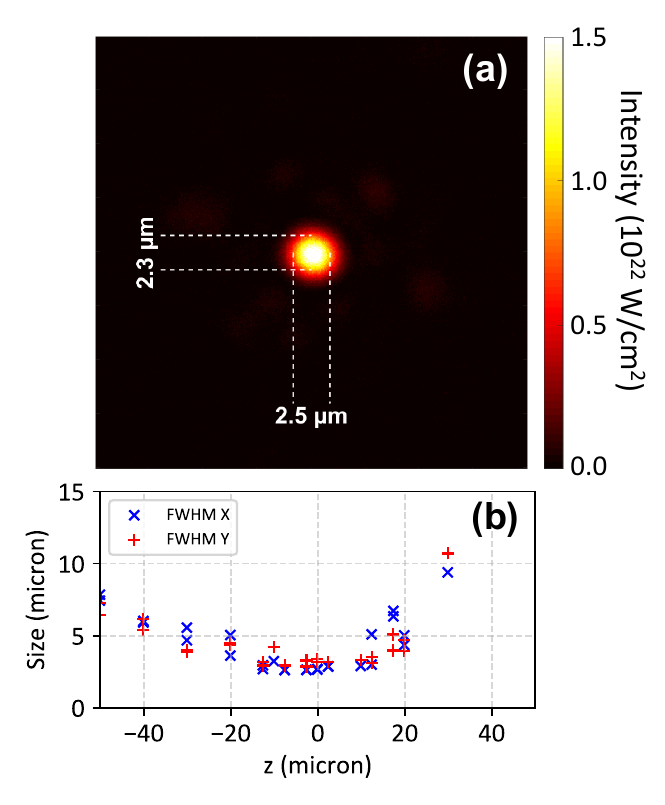}
    \caption{(a) Far-field measurement of the laser intensity distribution. 
    (b) Far-field beam focal spot scan, along the laser axis (z). 
}
    \label{fig:far-field}
\end{figure}

Far-field measurements of the laser intensity are shown in Fig.~\ref{fig:far-field}. From panel (a), the focal spot was measured to be slightly elliptical with $2.3\,\mu \mathrm{m}\times 2.5\,\mu \mathrm{m}$ FWHM dimensions, with about 44\% of the total energy within the first lobe of the beam and a corresponding Strehl ratio of about 51\%. The peak intensity, calculated from the spatial integration of the energy in each pixel, assuming that the full laser energy, i.e., 40 J, is contained in the image and taking into account the laser pulse duration, is $I_0 = 1.8\times10^{22}$ W/cm$^2$ (\textcolor{black}{which corresponds to $a_0 \equiv \sqrt{\frac{I_0\lambda^2\mu_0 q_e^2}{2 \pi^2 m_e^2 c^3}} \sim 90$, where $\mu_0$ is the permeability in vacuum, $q_e$, $m_e$, and $c$ are the elementary charge, the electron mass, and the speed of light, respectively}).
Panel (b) gives the results of the focal spot scan along the laser axis (i.e., $z$). When close to the target chamber center (TCC) ($|z| < 12\ \mu \mathrm{m}$), the size of the focal spot is quite stable around $2.5\ \mu \mathrm{m}$ in both the $x$ and $y$ dimensions. Further measurements are scheduled to allow a better estimation of the intensity capacity of Apollon, supported by spectrally resolved on-target wavefront acquisitions, taking into account residual spatiotemporal coupling effects. 

\begin{figure}[ht]
    \centering
    \includegraphics[width=0.45\textwidth]{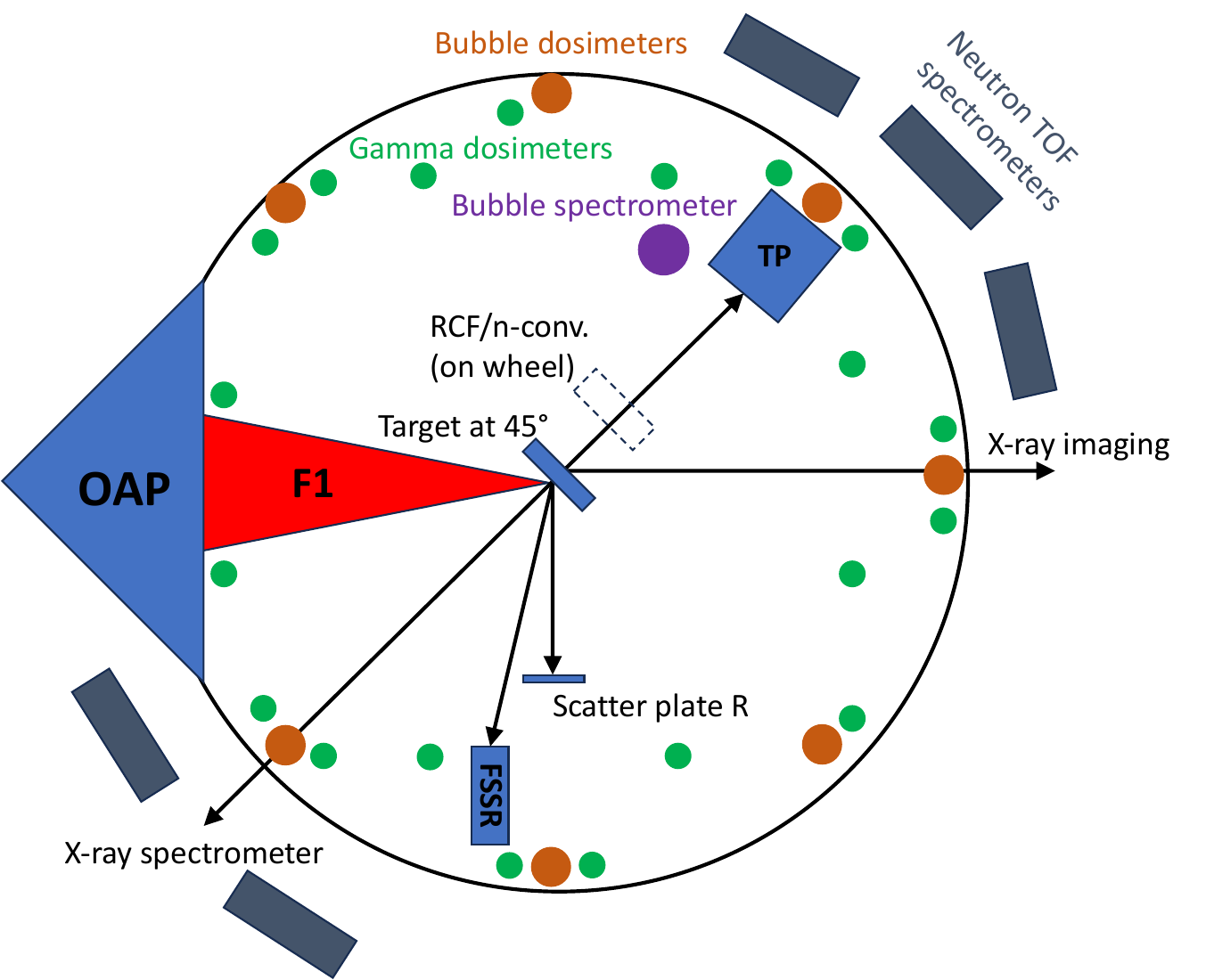}
    \caption{Schematic of the experimental setup. The diagnostics include the reflected laser beam (scatter plate R), plasma characterization (FSSR), accelerated ions (RCF and TP), X-rays imaging; together with diagnostics of the neutrons, i.e., nToF, which also allows for X-ray quantitative dosimetry, bubble dosimeters, and bubble spectrometer. }
    \label{fig:setup}
\end{figure}

\section{Experimental setup and characteristics of the plasma conditions}
\label{sec:setup}

\begin{figure}[ht]
    \centering
    \includegraphics[width=0.3\textwidth]{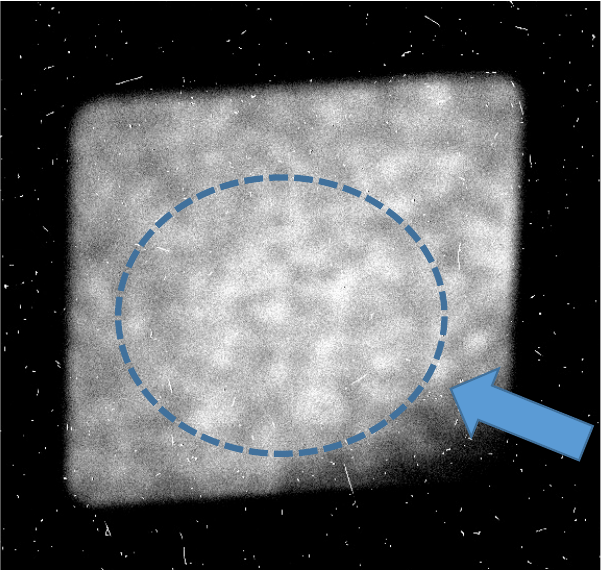}
    \caption{Image of the specularly reflected laser beam landing on the Spectralon plate after having irradiated a 6 $\mu$m thick Al target, as captured by an imaging camera positioned outside the chamber. The blue arrow indicates the direction of the reflected laser propagation and the dashed circle indicates the expected pattern of the laser if it would have been perfectly reflected by the target.}
    \label{fig:spectralon}
\end{figure}

The experimental setup is illustrated in Fig.~\ref{fig:setup}. The beam was focused onto the targets with a 45\degree  \ incidence angle. Aluminum foils of $1.5$--$15\,\mu \rm m$ thicknesses were used. As for the diagnostics, along the specular reflection axis, a camera recorded the image of the reflected laser beam on a Spectralon (i.e., Lambertian) scatter plate, as shown in Fig.~\ref{fig:spectralon}. Such image clearly shows that the laser beam reflected from the target was widely divergent and scattered.
Such a pattern indicates a poor temporal contrast, inducing a long preplasma at the target front, in which the incident laser filaments, yielding the observed reflected beam pattern (see the comparison between Fig. 6 (c) and (d) of Ref. \onlinecite{burdonov2021characterization}).

\begin{figure}[ht]
    \centering
    \includegraphics[width=0.48\textwidth]{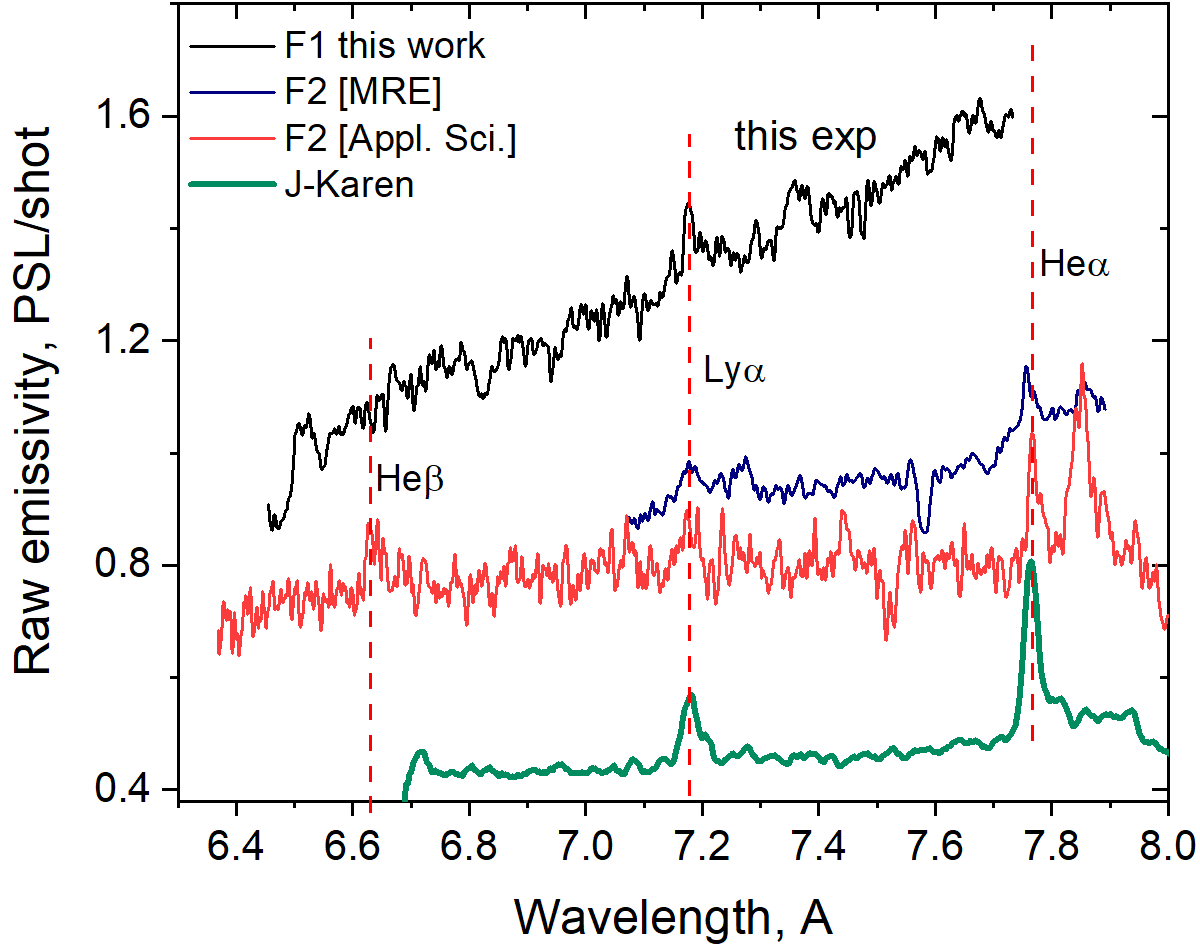}
    \caption{X-ray spectra of Al ion emission measured with the FSSR spectrometer equipped with a mica crystal in different experimental campaigns -- at the Apollon facility (black curve) using the F1 [this work], or F2 (dark blue and red curves, respectively) \cite{burdonov2021characterization,yao2024enhanced} beamline, and at the J-KAREN laser facility~\cite{faenov2015nonlinear} (green curve). The detection parameters (i.e. filtering, TCC-crystal distance) were slightly different. All spectra measured in the cited papers were shifted in amplitude to compare with the current results for visibility. The dashed lines represent positions of spectral components: Al Ly$\alpha$, He$\beta$ and He$\gamma$, respectively.
}
    \label{fig:fssr}
\end{figure}

The presence of such a large pre-plasma at the target front is further corroborated by a spatially resolved X-ray spectrometer (FSSR) \cite{Faenov1994} that was fielded beside the scatter plate. The result of the FSSR measurements is shown in Fig.~\ref{fig:fssr}. The Al Ly$\alpha$ line was measured, implying that a quite high electron $T_e$ was achieved. However, we observed that the ratio between the X-ray resonance lines to Bremsstrahlung was much lower than in other experiments performed using fs-scale laser irradiating thin foils up to several micrometers \cite{faenov2015nonlinear, soloviev2017experimental,burdonov2021characterization}. This indicated here a more extended plasma source, possibly due to a low laser contrast. Quantitatively, we could not compare the present results to those former works, because the experimental range of the FSSR in the F1 campaign was limited (due to the mechanical issues) and so there was no information about the He$\alpha$ emission which should be dominant. 

Further, we characterized the bulk temperature from x-ray spectra. For this, we used solid multi-layered foils composed of a 0.2 $\mu$m-thick titanium layer between two 2 $\mu$m-thick
plastic layers. These plastic layers serve as tampers to limit the fast expansion of the titanium plasma.
They also prevent a direct interaction between the laser and the titanium. This titanium layer serves as a tracer: its K-shell x-ray emission (from 4.5 to 6 keV) depends strongly on the electron temperature.

To measure the relevant x-ray spectra, a poly-capillary x-ray optics \cite{dorchies2015experimental} was
used to collimate, from the target front (see Fig.~\ref{fig:setup}), the x-rays and send them to an auxiliary vacuum chamber 2 meters away from the interaction.
This technique has major advantages in the context of ultra-high-intensity laser-plasma interaction, as the poly-capillary brings the signal far from the extreme environment in the vicinity of the interaction.
At that distance, electronic systems survive the shot, and the x-ray CCD cameras capture very little of the noise that is generated in the main vacuum chamber.
As a result, we were able to obtain x-ray spectra with an excellent signal-to-noise ratio.

The collimated x-rays were sent on a curved Bragg crystal for dispersion. It consists of a curved layer of pentaerythritol mounted on an aluminum substrate.
The polynomial curvature of the crystal is designed to maintain a high spectral resolution at a given plane a few centimeters away, where a CCD camera is placed.

\begin{figure}[ht]
    \centering
    \includegraphics[width=0.48\textwidth]{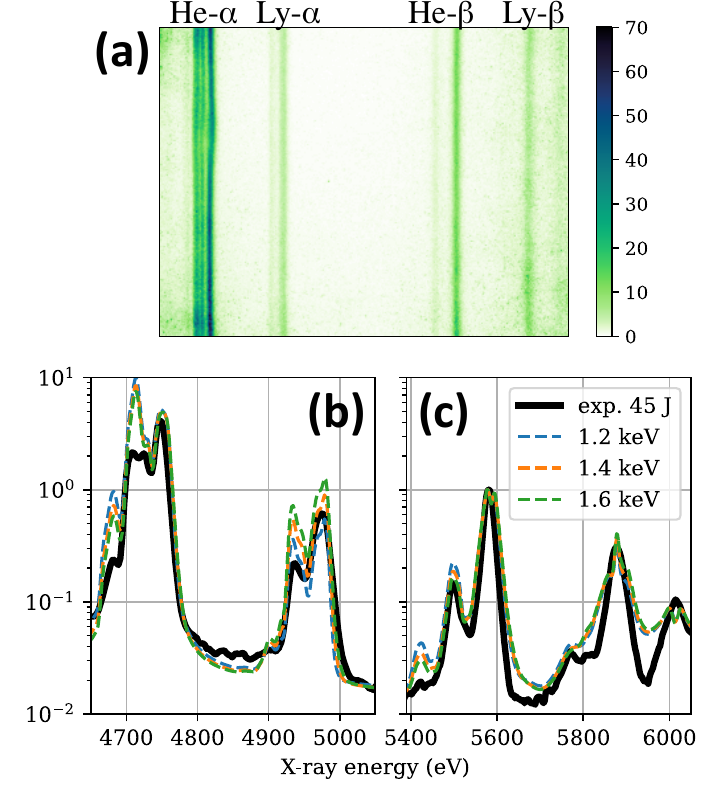}
    \caption{(a) Typical Ti x-ray spectrum detected on a CCD camera, using a poly-capillary x-ray optics and a curved Bragg crystal. The dispersion axis is horizontal. (b) and (c) Experimental (solid line) and simulated (dashed lines) Ti x-ray spectra. Simulations assumed a density of 2 g/cm$^3$. All curves have been normalized to the peak value of the He-$\beta$ line at 5580 eV.}
    \label{fig:heating}
\end{figure}

An example of a spectrum is provided in Fig.~\ref{fig:heating} (a), illustrating the typical noise level and the intensity of the various x-ray lines. To obtain this spectrum, some processing was carried out. Indeed, the background level of the camera and the response of the x-ray optics must be taken into account.
First, the background level is obtained by subtracting the average number of counts measured in a portion of the CCD chip where no signal from the Bragg crystal can be received. Second, the response of the x-ray system is inferred from separate ``reference'' shots on targets that are known to provide a nearly flat spectrum.
In particular, we used the signal obtained from shots on aluminum targets.
Dividing the raw signal by this reference data gives a spectrum corrected for the response of all optical elements, including filters.
Note that, in that energy range, aluminum emits a spectrum that is not perfectly flat, but slowly decreasing with the photon energy. It was accounted for by simulating this theoretical spectrum using atomic physics codes.
This correction however mildly impacts  the results.

To estimate the target bulk temperature from these spectra,
we employed similar atomic simulations provided by the code \textsc{SAPHyR}, which is based on a collisional-radiative model to describe atomic, kinetic and spectral properties of a wide variety of non-local-thermodynamic-equilibrium plasmas including mixtures.
Many simulations of the Ti plasma were carried out with different values of density and temperature, resulting in numerical spectra that were compared to the experimental data.
In a first approach, we assumed that the density of the titanium was uniform, but that its temperature had a Gaussian transverse profile. We also neglected temporal variations. This basic analysis cannot be expected to reproduce spectra with great accuracy, but it provides an estimate of the ratios between the main x-ray lines that are sensitive to the temperature, especially between the helium-like and hydrogen-like lines.

Figure~\ref{fig:heating} (b) and (c) show an example of comparisons between experimental and simulated spectra.
Significant differences can be observed close to the He-$\alpha$ line (4750 eV), but which is known to have strong opacities, and for which the calculation of the self-absorption may not be adequate in these circumstances. Our focus is mainly on the Ly-$\alpha$ (4900-5000 eV) and He-$\beta$ (5500-5600 eV) lines, which are well reproduced.
Matching their ratio leads to an average electron bulk temperature close to 1.4 keV. This typical result shows that the laser energy couples well to the target, reaching high temperatures in its center. Additional analysis will be carried out (with a combination of radiative-hydrodynamic and atomic modeling) to characterize the emission spectra accurately and to compare how it matches with the heating calculated by laser-plasma interaction simulations.

\section{Characterization of proton generation}
\label{protons}

Along the target rear normal axis (see Fig.~\ref{fig:setup}), the spectra of the emitted proton  were measured using stacks of HDv2 and EBT3 Gafchromic radiochromic films (RCFs) \cite{chen2016absolute} interlaced with aluminum filters, placed at 31~mm from the target on wheel. 

\begin{figure}[ht]
    \centering
    \includegraphics[width=0.4\textwidth]{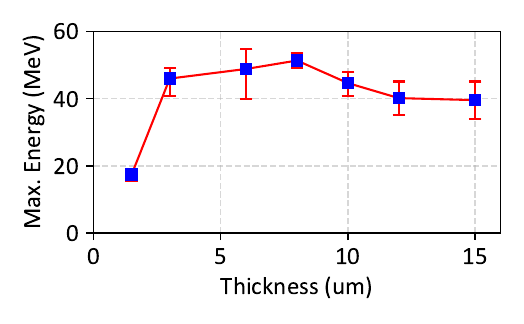}
    \caption{Proton cutoff energy as a function of the target thickness, as measured by RCFs \cite{bolton2014instrumentation}. Error bars on the proton energy represent the shot-to-shot fluctuation, accumulated with the simulated measurement uncertainty in the detector.
    }
    \label{fig:rcf}
\end{figure}

Figure~\ref{fig:rcf} shows the variation of the observed proton cutoff energy inferred from these RCF stacks, as a function of the target thickness. We find that the optimum target thickness is around 6 to 8 $\mu$m, which is also consistent with a poor laser contrast since thinner targets were decompressed prior to the main laser interaction \cite{fuchs2006laser,yao2024enhanced}. With an average on-target laser energy of 40 J, we can get stably around 50 MeV protons. 

\begin{figure}[ht]
    \centering
    \includegraphics[width=0.4\textwidth]{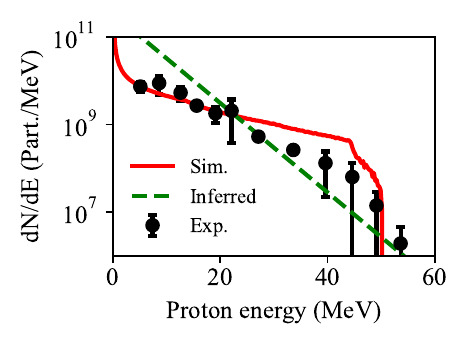}
    \caption{Proton energy spectrum, averaged from the RCFs data over 3 different shots (black dots), and compared with 2D PIC simulation results (red solid line).
    Overlaid is the steeper proton spectrum inferred from the neutron measurements (green dashed line).
    }
    \label{fig:spec_rcf}
\end{figure}

The proton energy spectrum averaged over 3 shots and extracted from the RCFs \cite{bolton2014instrumentation} is shown in Fig.~\ref{fig:spec_rcf}. 
Overlaid in Fig.~\ref{fig:spec_rcf} (see the green dashed line) is another spectrum, which was found to fit better the neutron emission measurement (as will be detailed below in Sec.~\ref{neutron}). Since the neutron generation and measurements were performed in a later phase of the experiment, we can deduce that the RCF-measured and neutron-inferred spectra represent the range of proton spectra that were actually generated during the beamtime. 

In addition, the red solid line in Fig.~\ref{fig:spec_rcf} shows the corresponding particle-in-cell (PIC) simulation result. 
The simulations are performed in two dimensions (2D), using the fully-relativistic \textsc{SMILEI} PIC code \cite{derouillat2018smilei}. The simulated targets are 6 $\mu$m in thickness, and are coated with a 20 nm thick hydrogen contaminant layer on their backside.
As detailed in Ref.~\onlinecite{horny2024lighthouse}, we used the radiative hydrodynamic simulation code MULTI to calculate the pre-plasma profile (that is initialized in the \textsc{SMILEI} simulation).  The pre-plasma results from a laser prepulse of 2 ps at the level of 10$^{15}$ W/cm${^2}$, located 100 ps before the main pulse. 
The intensities used in the \textsc{SMILEI} simulations are lower than the experimental ones, due to 2D simulations overestimating the interaction \cite{liu2013three,yao2024enhanced}. We thus adjusted the laser intensity to match the measured proton spectrum (see Fig.~\ref{fig:spec_rcf}). 

Note that in our previous campaign at Apollon, using the secondary F2 beam with about 10 J energy \cite{yao2024enhanced}, we measured on average a proton cutoff energy of about 25 MeV. Here, with the F1 beam at 40 J, our best proton cutoff energy is about 50 MeV, which fits quite well a trend following the scaling factor of $E^{0.5}$ \cite{fuchs2006laser}, i.e., with a four-times higher laser energy, the cutoff energy is doubled. 
The laser-to-proton energy conversion for this campaign, evaluated by integrating (over $0 \le E \le E_{\rm max}$) the exponential fit of each recorded proton spectrum, as inferred from the RCF-measured spectra, is estimated to be between 0.7\% and 0.9\%, i.e. lower than in the former campaign with the F2 beam \cite{yao2024enhanced}. This is understandable as in general high-contrast laser-target interactions lead to a better coupling efficiency compared to low contrast interactions \cite{yao2024enhanced}, where part of the laser is dissipated in the long preplasma at the target front.

\begin{figure}[ht]
    \centering
    \includegraphics[width=0.4\textwidth]{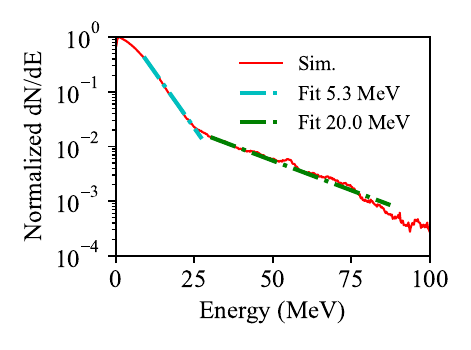}
    \caption{Simulated electron energy spectrum, along with the fitted electron temperatures. 
    }
    \label{fig:spec_elec}
\end{figure}

\begin{figure*}[ht]
    \centering
    \includegraphics[width=0.8\textwidth]{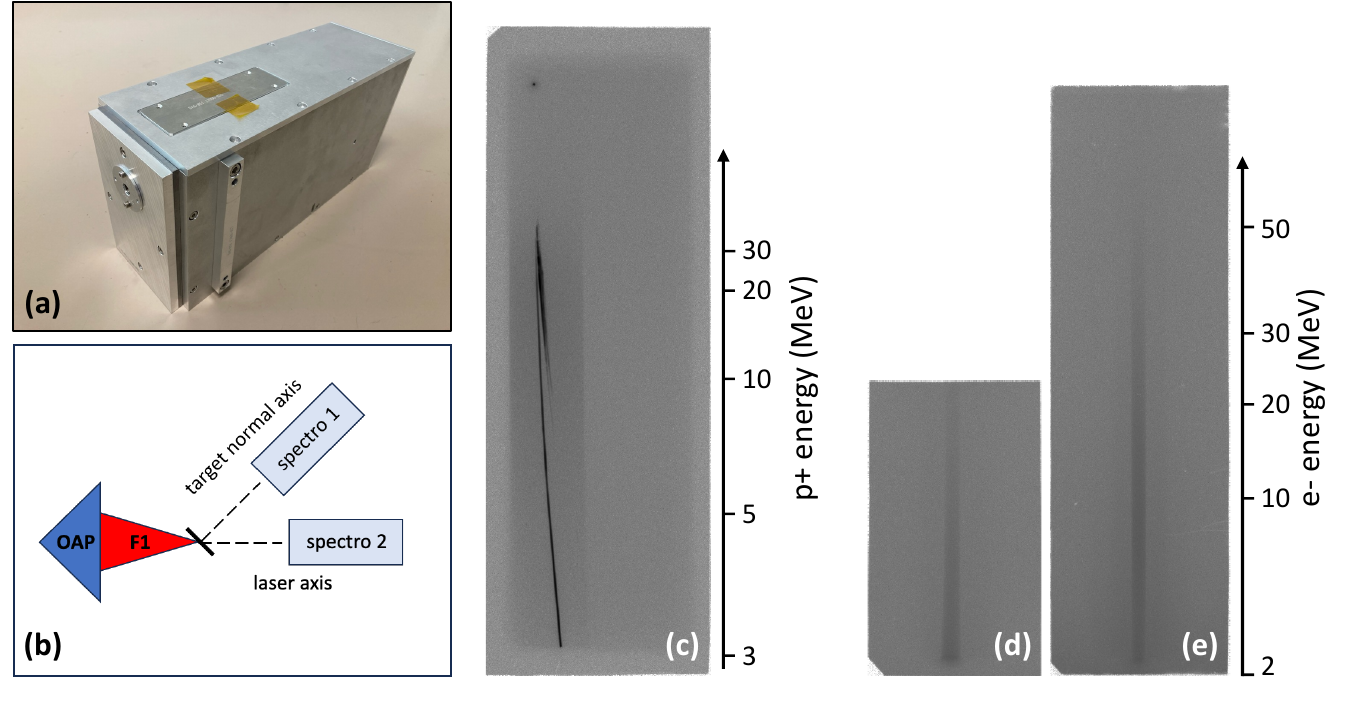}
    \caption{(a) Ion-electron spectrometer. (b) Experimental arrangement for the diagnostics tests. (c) Ion spectra from the irradiation of 10 $\mu$m thick Titanium target recorded on imaging plates along the target normal direction. (d) Electron spectra from the irradiation of layered CH-Ti-CH targets recorded on imaging plates along the target normal direction and (e) along the laser axis direction.
    }
    \label{fig:ie-spec}
\end{figure*}

The electron energy spectrum, as given by the simulation (adjusted to match the experimental proton spectrum), is shown in Fig.~\ref{fig:spec_elec}. The spectrum shows a typical two-temperature distribution. Maxwellian fittings give a hot electron temperature of 5.3 MeV for energies lower than 25 MeV, \textcolor{black}{in good agreement with the $(I_0\lambda^2)^{1/3}$ scaling \cite{beg1997study} and the relativistic model \cite{haines2009hot}.
However, for the energy range higher than 25 MeV, the hot electron temperature, fitted to be 20 MeV, is closer to the ponderomotive scaling \cite{wilks1992absorption}, pointing to different acceleration mechanisms in different parts of the target (i.e. the long preplasma at the target front and the dense target, respectively). }

Part of the commissioning campaign was also dedicated to the testing of compact, integrated ion-electron spectrometers. These spectrometers exploit an ad-hoc design to simultaneously record the spectra of electrons and of the various ion species emitted from a laser-irradiated target. In the spectrometers all charged particles are dispersed spectrally by a 1 T static magnetic field and different ion species are additionally separated by a parallel MV/m electric field. The spectrometers thus operate as magnetic spectrometers for the electrons and as Thomson spectrometers \cite{Thomson1907}  for the ions. A specifically developed design allows a partial compensation for the different energy deposition properties of ions and electrons, which normally would result in a much lower electron than ion signal (more details will be provided in a dedicated article). The spectrometers have been designed to be compact and easy to displace inside the experimental chamber, measuring between 25 and 30 cm in length depending on the specific model, and can be used in an open configuration with the ion detector located at some distance outside the spectrometer to increase spatial dispersion. For the tests, two spectrometers were simultaneously installed in the Apollon vacuum chamber, one along the target normal (where ion signal is maximized) and one along the laser axis (where electron signal is maximized) at about 55 cm from the target. Metal and plastic foils of typical few $\mu$m thicknesses were used as targets. To detect the charged particles, the spectrometers can be equipped either with radiation sensitive films (e.g., imaging plates \cite{bolton2014instrumentation} or RCF), scintillators or active detectors (microchannel plates \cite{bolton2014instrumentation}, C-MOS detectors \cite{Burdonov2023}, etc). Typical examples of ion and electron spectra recorded on imaging plates are shown in Fig.~\ref{fig:ie-spec}. Fig.~\ref{fig:ie-spec} (c) shows ion spectra recorded on a single shot, while Fig.~\ref{fig:ie-spec} (d)-(e) show electron spectra recorded by accumulating the signal over 5 shots.
Part of the commissioning was dedicated to the comparison, in terms of sensitivity and overall efficiency, of various types of plastic (Eljen EJ-208 and EJ-260, Saint-Gobain BC-430) and phosphor (P43 type) scintillators as ion and electron detectors for the spectrometers. The use of scintillators, with CCD cameras to record the scintillation signal located outside the experimental chamber, is in fact made necessary in experiments were the level of electromagnetic noise (EMP) or energetic radiation is too high to operate, inside the chamber, detectors with integrated electronics \cite{burdonov2021characterization}. This was indeed an issue that manifested with the use of C-MOS detectors during the commissioning campaign. Comparison of different types of scintillators as particle detectors is ongoing in the frame of a collaboration with the Rutherford Appleton Laboratory, and more details will be provided in a dedicated article.

\begin{figure}[ht]
    \centering
    \includegraphics[width=0.48\textwidth]{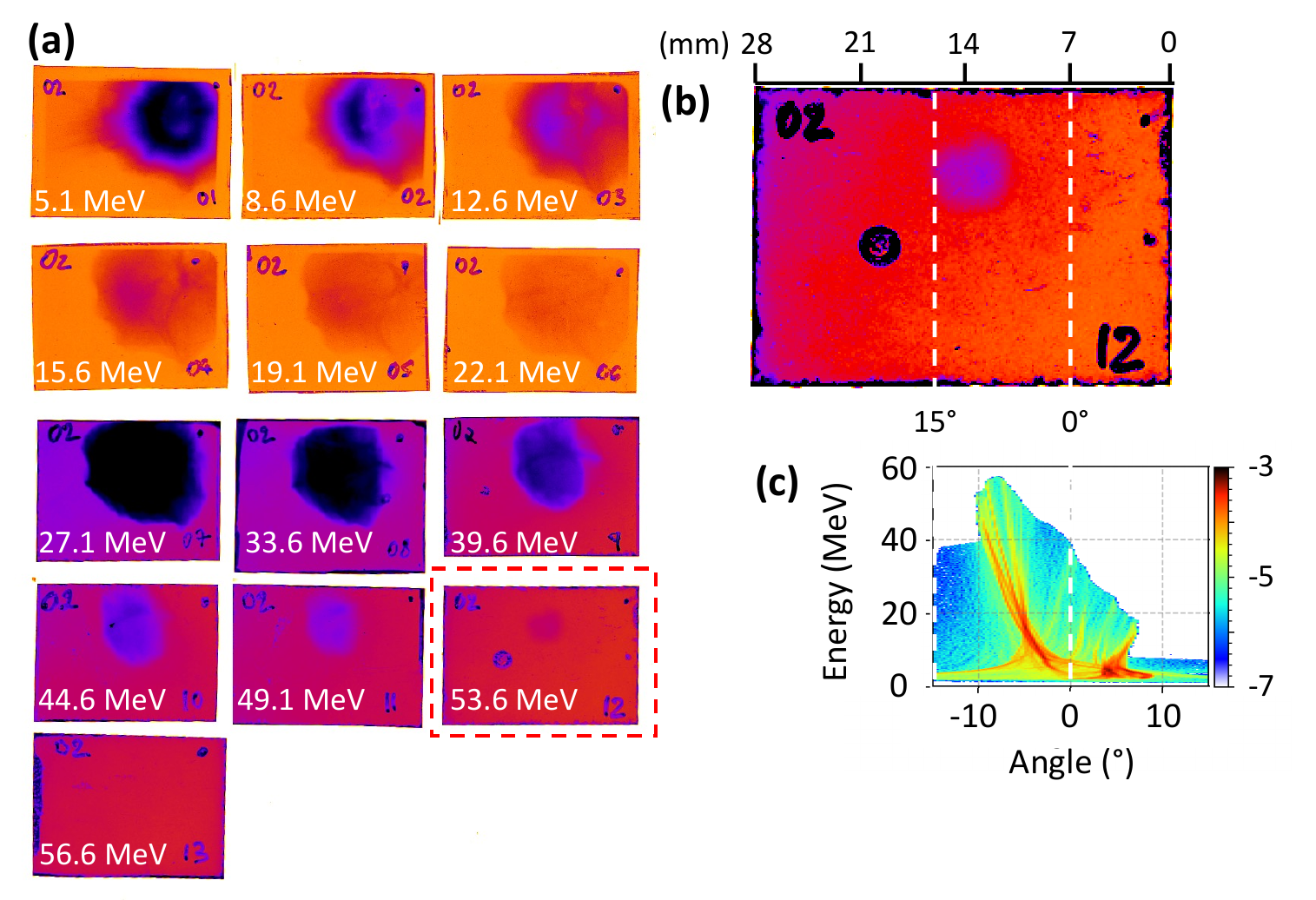}
    \caption{(a) Raw RCF data of shot 33. The first six layers are HDv2 films, while the other layers are EBT3 ones. The strong difference in sensitivity of these two films \cite{chen2016absolute} explains that the proton imprint is weak in the 6th layer, while it becomes strong in the 7th  one. (b) Zoom on the 12th layer in the red dashed box of (a), with adjustment of the image contrast for better readability. (c) The result of the 2D PIC simulations, which is qualitatively consistent with our measurement. 
    The angle here is defined as $\arctan(p_y/p_x)$, where $p_x$ and $p_y$ are the longitudinal and transverse momentum of the protons. 
    }
    \label{fig:RCF_angular}
\end{figure}

Figure~\ref{fig:RCF_angular} (a) shows the raw RCF images for protons accelerated from an Al target with a thickness of 6 $\mu$m, with the proton energy corresponding to each layer marked on the left bottom corner. 
As shown in Fig.~\ref{fig:RCF_angular} (b), on the zoom of the last (i.e., 12th) RCF layer, the high-energy beam profile can be seen clearly to be shifted away from the target normal axis (marked as the 0$\degree$ axis) by about 15$\degree$. 

The energy-angular distribution of the corresponding PIC simulation result is shown in Fig.~\ref{fig:RCF_angular} (c). It is clear that for the high-energy part of the proton beam, the angular distribution is shifted away from the target normal (0$\degree$) towards the laser axis, also within 15$\degree$, similar to what is shown by the RCF measurements. We have already observed such shift in our former experiments conducted using the F2 beam at Apollon, which we attributed as due to the pre-pulse pre-expanding the target \cite{horny2024lighthouse}.

\section{Characterization of neutron emissions}
\label{neutron}

Neutrons were generated from the protons using the standard pitcher-catcher scheme \cite{lancaster2004characterization,higginson2011production}. For this, a converter consisting of a 8 mm thick and 1 inch diameter lithium fluoride (LiF) disk was used. The neutrons were generated from the proton beams mainly through the $^7\rm Li(p,n)^7Be$ and $^{19}\rm F(p,n)^{19}Ne$ reactions. A 2 mm thick lead slab was added behind the converter to prevent the direct interaction of protons with energies higher than 56 MeV with the neutron diagnostics.

The neutron emissions were characterized using various methods \cite{lelievre2023comprehensive}. These include activation foils, bubble dosimeters and bubble spectrometer, from Bubble Technology Industries (BTI \cite{BTI}) inside the target chamber, while neutron time-of-flight (nToF) spectrometers were placed outside of the target chamber, both along the target normal axis and the specular reflection axis.

\subsection{Activation measurements}
\label{Activation}

The activation diagnostic was composed of an indium foil (thickness = 2 mm, diameter = 22 mm,  $\rm\rho = 7.30\,g/cm^3$), in which two main reactions can occur: $^{115}\rm In(n,\gamma)^{116m}In$ and $^{115}\rm In(n,n')^{115m}In$. The first reaction, producing $^{116m}\rm In$ nuclei ($\rm T_{1/2}=54.1 min$), is particularly sensitive to neutrons below 2 MeV (and mostly to epithermal neutrons, i.e. with energies in the 0.025-0.4 eV window, see  Fig. 3 of Ref.~\onlinecite{Lawriniang2015}), while the second reaction, producing $^{115m}\rm In$ nuclei ($\rm T_{1/2}=4.49 h$), has significant cross sections in a complementary window, i.e. between 2 and 10 MeV (see Fig. 9 of Ref.~\onlinecite{lelievre2023comprehensive}). 
It is followed by a magnesium foil (thickness = 10 mm, diameter = 22 mm, $\rm\rho = 1.74\,g/cm^3$), which has complementary significant cross sections between 6 and 30 MeV (see also Fig. 9 of Ref.~\onlinecite{lelievre2023comprehensive}) considering the $^{24}\rm Mg(n,p)^{24}Na$ reaction producing $^{24}\rm Na$ nuclei ($\rm T_{1/2}=14.96 h$).
This activation stack was placed at $0\degree$, directly behind the lead slab and the LiF converter to capture as many neutrons as possible and to maximize the activation of the foils.

The LiF converter and activation foils were measured by gamma spectrometry using a 3$\times$3" NaI scintillator, shielded by 10 cm of lead. The gamma-ray emission spectra were digitized with a Canberra OSPREY Multi-Channel Analyzer (MCA) and analyzed with the Genie2000 software, which allows for background noise subtraction and peak fitting. The efficiency for the determination of the foils activity in the present measurement geometry (at contact) was previously determined by reference gamma spectrometry of dissolved foils in a standard geometry of counting (50 mL vessel). \textsc{Geant4} simulations \cite{agostinelli2003geant4} were performed considering the estimated proton spectrum with a mean energy of 4.25 MeV, shown in Fig.~\ref{fig:spec_rcf} (dashed green line) and considering a conical beam for the protons \cite{bolton2014instrumentation,Snavely2000}, with a fixed divergence of $21\degree$ for all proton energies. These simulations indicate that the $^7\rm Li(p,n)^7Be$ reaction contributes to around 76.7\% to neutron production. The produced nuclei, $^{7}\rm Be$, can be easily measured by gamma spectrometry, unlike the $^{19}\rm Ne$ nuclei produced by the $^{19}\rm F(p,n)^{19}Ne$ reaction. Therefore, measuring the $^{7}\rm Be$ residual nuclei within the converter provides a reliable estimation of the total neutron yield.

The LiF converter, used during a series of 19 shots, exhibited a $^{7}\rm Be$ activity of 900.04 $\pm$ 50.80~Bq, corresponding to 47.37 $\pm$ 2.99 Bq/shot and a neutron production of
$(3.14\pm0.20)\times 10^{8}$ neutrons/shot from the Li nuclei. Given their 76.7\% contribution to the total neutron production, it can be deduced that approximately $(4.10\pm0.26)\times 10^{8}$ neutrons/shot were emitted during this series of shots, which aligns closely with the predicted value of $4.09\times 10^{8}$ neutrons/shot from the simulations.

The reactions of interest, and the activities of the foils used during the same series of 19 shots, are presented in Table.~\ref{tab_activation}, alongside the simulation results. These activities were measured at the end of the series of shots, they are therefore affected by the decay of radionuclides between shots. Thus, the measured activities were corrected by the mean decay time of 5.3 minutes between shots to obtain the average activities induced per shot.

\begin{table}[ht]
\begin{tabular}{ccccc}
\toprule
\toprule
Reaction & A$_{\rm sim}$/shot & A$_{\rm meas}$/shot & Measurement  \\
& (Bq) & (Bq) & uncertainty \\
\midrule
\addlinespace[0.1cm]
$^{115}\rm In(n,\gamma)^{116m}In$  & 17.49 & 9.92 & 13.40 \% \\ 
\addlinespace[0.2cm]
$^{115}\rm In(n,n')^{115m}In$  & 6.74 & 7.17 & 7.10 \% \\
\addlinespace[0.2cm]
$^{24}\rm Mg(n,p)^{24}Na$ & 1.31 & 1.02 & 9.16 \% \\
\addlinespace[0.1cm]
\bottomrule
\bottomrule
\end{tabular}
\caption{\justifying{Measured activities, A$_{meas}$, and their corresponding uncertainties. The simulated activities, A$_{sim}$, were computed with MCNP6 simulations using the expected neutron spectrum obtained from the proton spectrum guess, shown in Fig.~\ref{fig:spec_rcf}, and the IRDFF-II library \cite{TechReport_2023_LANL_LA-UR-22-33103Rev.1_RisingArmstrongEtAl}.}}
\label{tab_activation}
\end{table}

The measured activity of $^{116m}\rm In$, produced via the $^{115}\rm In(n,\gamma)^{116m}In$ reaction in the indium foil, points to a weaker low-energy neutron component in the actual spectrum, compared to that of the simulations. 
This could be due to the proton spectrum guess (green dashed line in Fig.~\ref{fig:spec_rcf}) overestimating the number of low-energy protons (see the saturation of the proton spectrum at low energy in the RCF-inferred spectrum shown in Fig.~\ref{fig:spec_rcf}).
However, the measured activities of $^{115m}\rm In$ and $^{24}\rm Na$, i.e., in a higher energy range, are close to the simulated values, suggesting that the simulations accurately reproduced the emissions of fast (MeV-range) neutrons.

Due to the relatively flat cross section profile of the $^{115}\rm In(n,n')^{115m}In$ reaction induced in the indium foil between 2 and 10 MeV, we can estimate the approximate number of neutrons emitted in this energy range using the equation:

\begin{equation}
N_{\Omega}=\frac{A_{\text {meas}}}{\lambda \times n_{\text {shot }}} \times \frac{1}{\bar{\sigma} \times t \times n \times \chi} \times \frac{1}{\Omega}
\end{equation}\\

\noindent Where $N_{\Omega}$ is the neutron fluence (neutrons/sr/shot), $A_{\text {meas}}$ the measured activity ($\mathrm{Bq}$), $\lambda$ the decay constant ($s^{-1}$), $n_{\text{shot}}$ the number of shots, $\bar{\sigma}$  the average cross section ($m^2$), $t$ the foil thickness (m), $n$ the number density ($m^{-3}$), $\chi$ the isotope abundance, and $\Omega$ the solid angle covered by the foil (sr).

Thus, a neutron fluence of $(1.97\pm0.14)\times 10^{7}$ neutrons/sr/shot between 2 and 10 MeV was obtained from the $^{115m}\rm In$ measured activity. This measured fluence is slightly lower but close to the simulated value of $2.34\times 10^{7}$ neutrons/sr/shot obtained in the same energy range (see Section~\ref{nToF}).

\subsection{Bubble detectors}
\label{Bubbles}

Two types of bubble detectors, supplied by \textit{Bubble Technology Industries} \cite{BTI}, were used: BD-PND dosimeters and a bubble detector spectrometer (BDS). BD-PND dosimeters are sensitive to neutrons from 200 keV to around 15 MeV, while the BDS consists of six types of bubble dosimeters with different sensitivity ranges (from 10 keV to 20 MeV), allowing to reconstruct neutron spectra in this energy range~\cite{ING1997}.

BD-PND dosimeters were positioned outside the interaction chamber, near the target normal axis ($0\degree$), at a distance of 130 cm from the converter. Due to their relatively low energy-dependent response between 200 keV and around 15 MeV, a neutron fluence of $(4.66\pm0.43)\times 10^{7}$ neutrons/sr/shot was obtained in this energy range from the measured doses, which is consistent with the simulated value of $5.06\times 10^{7}$ neutrons/sr/shot obtained in the same energy range (see Section~\ref{nToF}).

\begin{figure}[ht]
    \centering
    \includegraphics[width=0.48\textwidth]{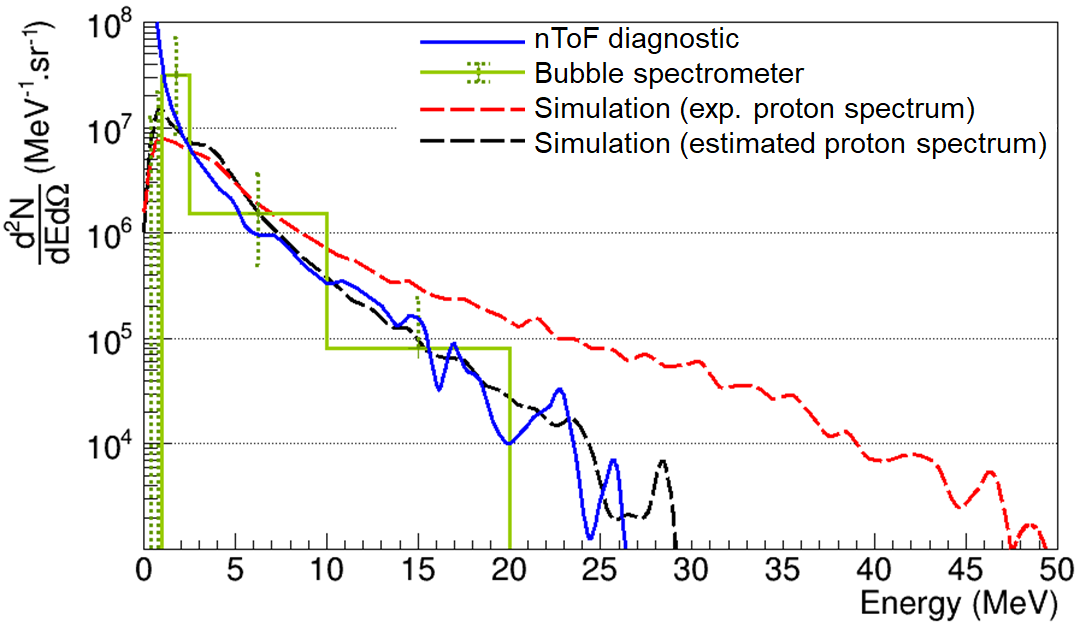}
    \caption{Comparison between the simulated neutron spectra (in red and black), obtained respectively from the experimental proton spectrum shown in Fig.~\ref{fig:spec_rcf} and the proton spectrum guess with a mean energy of 4.25 MeV, and the measured neutron spectra obtained with the bubble spectrometer and the nToF diagnostic.}
    \label{fig:Neutron_Comparison_Spectra}
\end{figure}

The BDS was placed near the target normal axis, at 60 cm from the converter. Fig.~\ref{fig:Neutron_Comparison_Spectra} shows a comparison of the simulated neutron spectra in the forward direction with the experimental spectra measured by the BDS and nToF diagnostic (as detailed below in Sec.~\ref{nToF}). The spectrum from the BDS closely matches, between 2 and 20 MeV, the nToF spectrum and the simulated spectrum obtained from the proton spectrum guess, corroborating the good agreement between the measured and simulated activities of $^{115m}\rm In$ and $^{24}\rm Na$ nuclei, as they are also sensitive in the same energy range. This alignment also supports the relevance of the proton spectrum estimate used to simulate neutron emissions, thus confirming the steeper slopes of proton spectra during shots with the LiF converter, compared to the RCF-measured one (see Fig.~\ref{fig:spec_rcf}).

However, as observed with the spectrum inferred from the nToF diagnostic, the BDS spectrum shows significantly higher emissions of low-energy neutrons compared to the simulation, contrasting with the $^{116m}\rm In$ nuclei measurement, which indicated fewer low-energy neutrons than expected. This discrepancy may stem from scattered neutrons. Indeed, the setup includes BDS and scintillators which occupy a substantial volume (over several hundreds of $\rm cm^{3}$) and induce considerable scattering. 

\subsection{nToF measurements}
\label{nToF}

Several ultra-fast organic scintillators, based on a plastic matrix of polyvinyl toluene, were also used during this experiment for nToF measurements. They were placed from $5\degree$ to $90\degree$ relative to the target normal axis and at distances from 5 m to 8 m. Each scintillator consisted of an 1-inch diameter and 40 cm long scintillator tube (EJ-254)~\cite{Eljen} connected to one photomultiplier tube (9112B)~\cite{ETEnterprises} on each side~\cite{lelasseux2021}. The signal was digitized at a sampling frequency of 500 MHz using a CAEN VX1730B digitizer~\cite{CAEN}.

Additionally, an innovative liquid scintillator-based detector was also used to measure the nToF spectra from a short distance, i.e., 3.2 m away from the chamber wall. Our intend here was to demonstrate that a compact nToF measurement is possible, despite the mixing of EMP noise and prompt gamma radiation with the neutron signals at such a short distance.
The detector consisted of BCF-91A wavelength shifting (WLS) fibers immersed in Ultima Gold XR liquid scintillator, packed in a 6 cm long, 2.6 cm inner diameter cylinder. Each WLS fiber was coupled, at the exit of the scintillator volume, to a 4 m long clear fiber \cite{kishon2019laser}.  The clear fibers were directly attached to a photomultiplier. This arrangement has two main advantages: the attenuation length of clear fibers is far larger than that of the WLS fiber and second, the immunity of the clear fiber to ionizing radiation is significantly superior in comparison to the WLS fibers.

\begin{figure}[ht]
    \centering
    \includegraphics[width=0.48\textwidth]{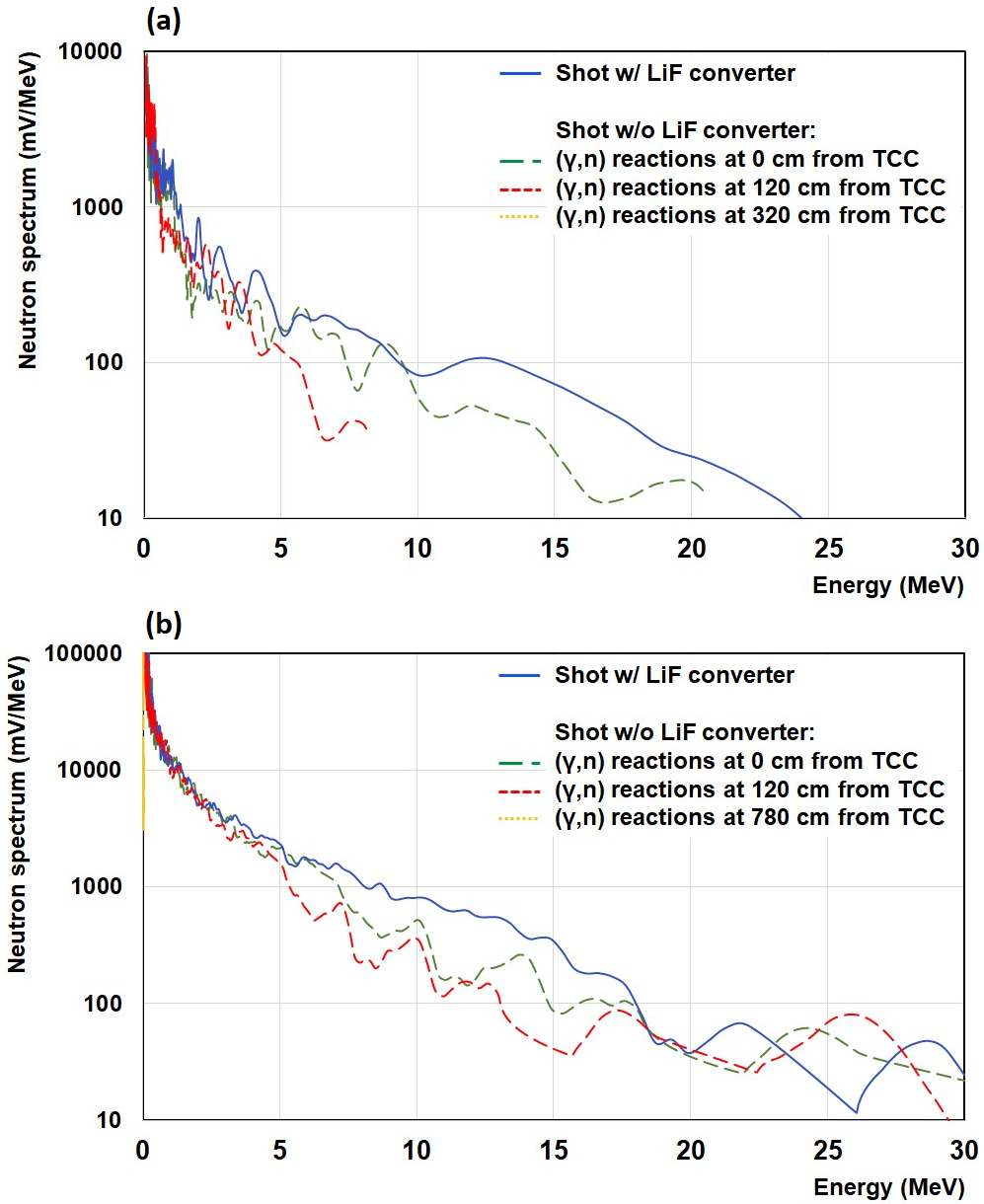}
    \caption{Neutron signals as a function of neutron energy, (a) using a liquid scintillator and (b) using a plastic scintillator, for shots with (blue solid lines) and without (dashed lines) LiF converter. For the shot without converter, different positions where ($\gamma$,n) reactions could occur were considered (i.e. at the location of the converter at the target chamber center (TCC), at the chamber wall (120 cm away) and at the location of the detector (320 cm away for panel (a) and 780 cm away for panel (b))). The red dashed lines and yellow dashed lines correspond, respectively, to photoneutrons produced in the chamber walls and in the lead shields.}
    \label{fig:nToF_signals}
\end{figure}

Several measurements, with and without a LiF converter were taken and analyzed. The solid blue line in Fig.~\ref{fig:nToF_signals} (a) shows the measured neutron spectra using the liquid scintillator detector. 
Fig.~\ref{fig:nToF_signals} (b) presents the same analysis, but for the nToF based on plastic scintillators, positioned 8 m away from the TCC and at $13\degree$ from the target normal axis. Due to the significant background noise induced by high X-ray emissions (see Section~\ref{x-ray_radio}), only this EJ-254 scintillator measured clear nToF signals, as it was furthest from the converter and shielded with a substantial amount of lead (15 cm thick lead bricks on the front face and 5 cm thick lead bricks on the top, bottom and rear faces).

Shots without converter are also shown in both Fig.~\ref{fig:nToF_signals} (a) and (b). The different curves represent the relationship between the neutron signal and neutron energy, considering various distances from the TCC where ($\gamma$,n) reactions may occur. It can be observed that the location where these photoneutrons are presumed to be produced significantly influences the energy derived from their ToF and, consequently, the shape and interpretation of the spectrum. For ($\gamma$,n) reactions assumed to occur near the TCC or within the chamber walls, the neutron spectra obtained are relatively similar to those measured when the LiF converter was used. This would suggest that neutrons are mainly produced via ($\gamma$,n) reactions, resulting in comparable neutron production with and without the converter. However, this assumption is inconsistent with results obtained from the bubble detectors and activation diagnostic shown above, which reveal no significant neutron emissions without the converter. Hence, these photoneutrons are more likely produced in the lead shields surrounding the scintillators (i.e. corresponding to the yellow curves in Fig.~\ref{fig:nToF_signals}), where the maximum cross section is 40 times higher than that of the aluminum composing the interaction chamber.

An unfolding procedure described in Ref.~\onlinecite{lelievre2023comprehensive} was applied to retrieve neutron spectra from the nToF signals obtained with the EJ-254 scintillator. The resulting experimental nToF spectrum is shown in Fig.~\ref{fig:Neutron_Comparison_Spectra}. It compares well  with the neutron spectrum obtained from the bubble spectrometer.

\begin{figure}[ht]
    \centering
    \includegraphics[width=0.48\textwidth]{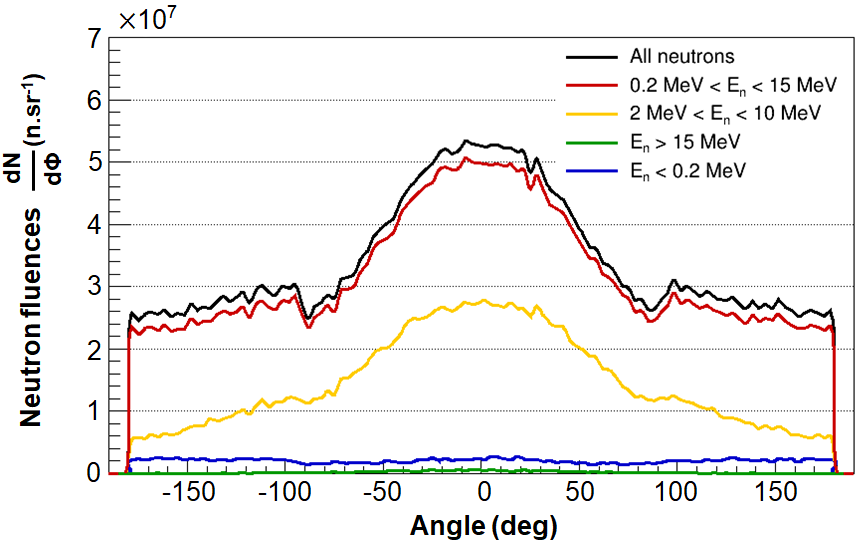}
    \caption{Simulated angular distribution of neutron emissions in the equatorial plane. The different curves represent the angular distribution in different energy ranges. The red and yellow curves correspond specifically to the angular distribution in the sensitivity range of the BD-PND dosimeters and the indium foil, respectively.}
    \label{fig:Neutron_Simulated_Angular_Distribution}
\end{figure}

The simulated angular distribution of neutron emissions across different energy ranges, shown in Fig.~\ref{fig:Neutron_Simulated_Angular_Distribution}, illustrates a clear anisotropy of emissions, particularly for neutron above 200 keV. A maximum neutron fluence of $5.33\times 10^{7}$ neutrons/sr/shot was obtained, with a fluence of $5.06\times 10^{7}$ neutrons/sr/shot in the sensitivity range of the BD-PND dosimeters, represented by the red curve. This confirms that their sensitivity range encompasses nearly all neutron emissions, providing reliable estimates of the emitted neutron fluences. Finally, a mean fluence of $2.34\times 10^{7}$ neutrons/sr/shot was expected in the solid angle covered by the indium foil and within its sensitivity range, i.e. between 2 MeV and 10 MeV. 

\section{X-ray generation and radiography capability}
\label{x-ray_radio}

\begin{figure}[ht]
    \centering
    \includegraphics[width=0.45\textwidth]{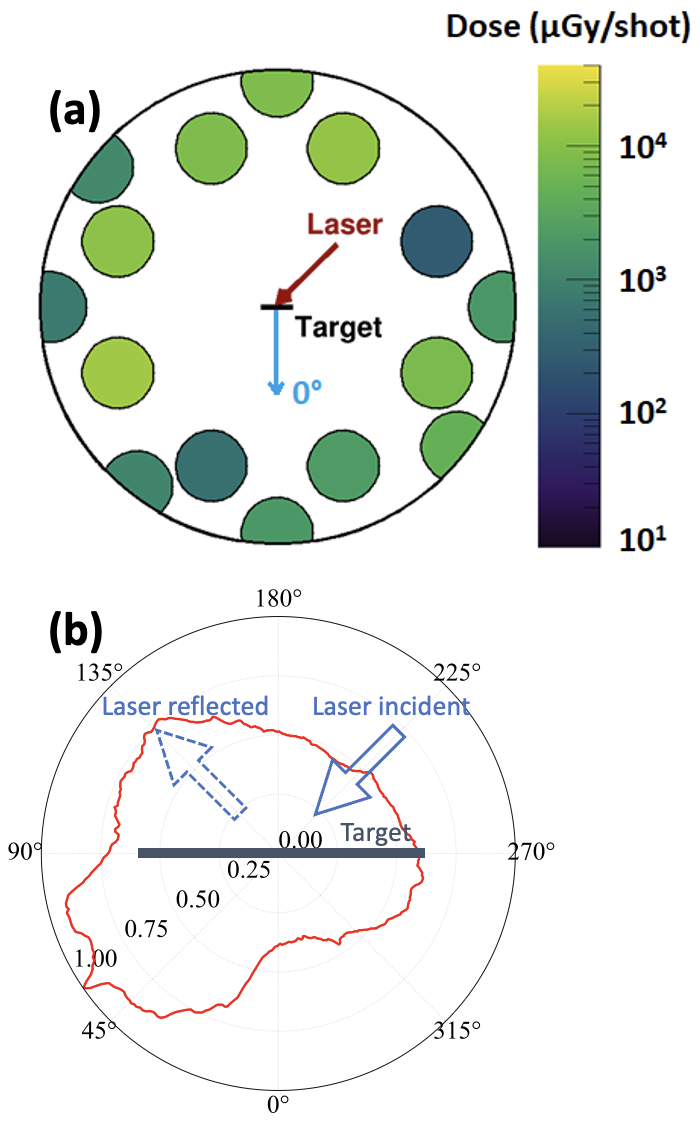}
    \caption{(a) X-ray spatial dose distribution measured inside the interaction chamber, using RPL dosimeters and during shots on 6 $\mu m$ and 8 $\mu m$ thick Al targets. Each circle represents the position of a dosimeter, and its color corresponds to the measured dose. (b) The corresponding PIC simulated EM-field induced X-ray energy distribution.}
    \label{fig:Dose_Map_RPL}
\end{figure}

The angular distribution of the X-ray emission produced from the laser-target interactions was measured using GD-351 Radio-Photo-Luminescent (RPL) dosimeters~\cite{WEIHAI2007}. These are positioned inside the interaction chamber, all around the target in the equatorial plane and at distances ranging from 87 cm to 115 cm. The dosimeters were encapsulated in a 1.5 mm thick tin holder, making them sensitive to X-rays in an energy range from around 30 keV to a few MeV, with an energy-independent response. A dose map, obtained from measurements accumulated over a series of 31 shots, is shown in Fig.~\ref{fig:Dose_Map_RPL} (a). These measurements reveal that X-ray emissions are stronger in the half-sphere that contains the laser transmission and the specular reflection axes, with an average maximum dose of 15.5 mGy/shot at 87 cm from TCC, in the transmitted direction of the laser. Note that the X-ray doses obtained during this experiment were around 10 times higher than those measured when the secondary F2 beam of the Apollon facility was used~\cite{yao2024enhanced}, although the presently used F1 beam has only four times more energy than the F2 beam. 

\begin{figure}[ht]
    \centering
    \includegraphics[width=0.42\textwidth]{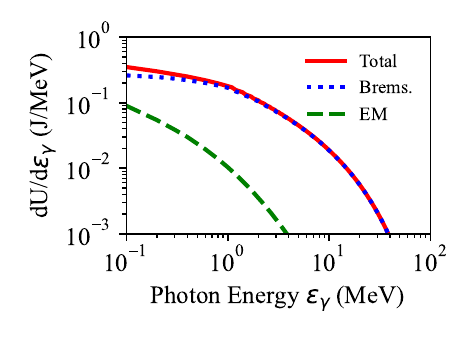}
    \caption{Simulated energy spectra of photons, both induced by EM-fields (green dashed line) and also from Bremsstrahlung (blue dotted line). See text for details.
    }
    \label{fig:spec_x}
\end{figure}

Since our diagnostic did not measure the X-ray energy distribution in the MeV-range, we performed theoretical estimates and numerical simulations of the X-ray generation to derive the X-ray spectrum. For this, we proceeded in two steps: 
(i) First, we have evaluated the electromagnetic (EM) field induced radiation, using the relevant module \cite{niel2018quantum} in the fully-relativistic \textsc{SMILEI} PIC code, which we already used to simulate the proton dynamics, see Fig.~\ref{fig:spec_rcf}. 
The resulting EM-field induced radiation is shown in Fig.~\ref{fig:spec_x} with a green dashed line.

(ii) Second, to estimate the Bremsstrahlung radiation we used the electron spectra from the \textsc{SMILEI} simulation (see Fig.~\ref{fig:spec_elec}) and estimated the energy loss by the refluxing of the electrons during the cooling of the plasma according to a target bulk heating model \cite{antici2013modeling}. The target bulk heating model was used to calculate the evolution of the electron temperature in time and the density drop of the plasma due to the expansion of the target. The Bremsstrahlung emission was then calculated similarly to \cite{myatt2009optimizing} with the modification that not all the electron energy is lost and that the ion density decreases over time. The electron energy loss is given from the target bulk heating simulation. 
Therefore, the Bremsstrahlung spectrum is given by:
\begin{equation}
N(k)=\eta_r N_e \int_0^{\infty}dT_0 f(T_0)  \int ^{T_0}_{T_{end}}dT n_i \sigma_{\gamma}(E,k)dk |\frac{dT}{ds}|^{-1}
\end{equation}
where $\eta_r$ is the refluxing efficiency, $N_e$ is the total number of the hot electrons, $f$ is the normalized electron distribution, $T_{end}$ is the final kinetic energy of the electron at the end of the target bulk heating simulation, $n_i$  is the ion density of the target, $\sigma_{\gamma}$ is the bremsstrahlung cross-section, $k$ is the photon energy, $E$ is the total electron energy, and $\frac{dT}{ds}$ is the energy loss of the electron.
The resulting Bremsstrahlung spectrum is shown in Fig.~\ref{fig:spec_x} with a blue dotted line.
It is clear that the radiated X-rays are here dominated by Bremsstrahlung. 

\begin{figure}[ht]
    \centering
    \includegraphics[width=0.42\textwidth]{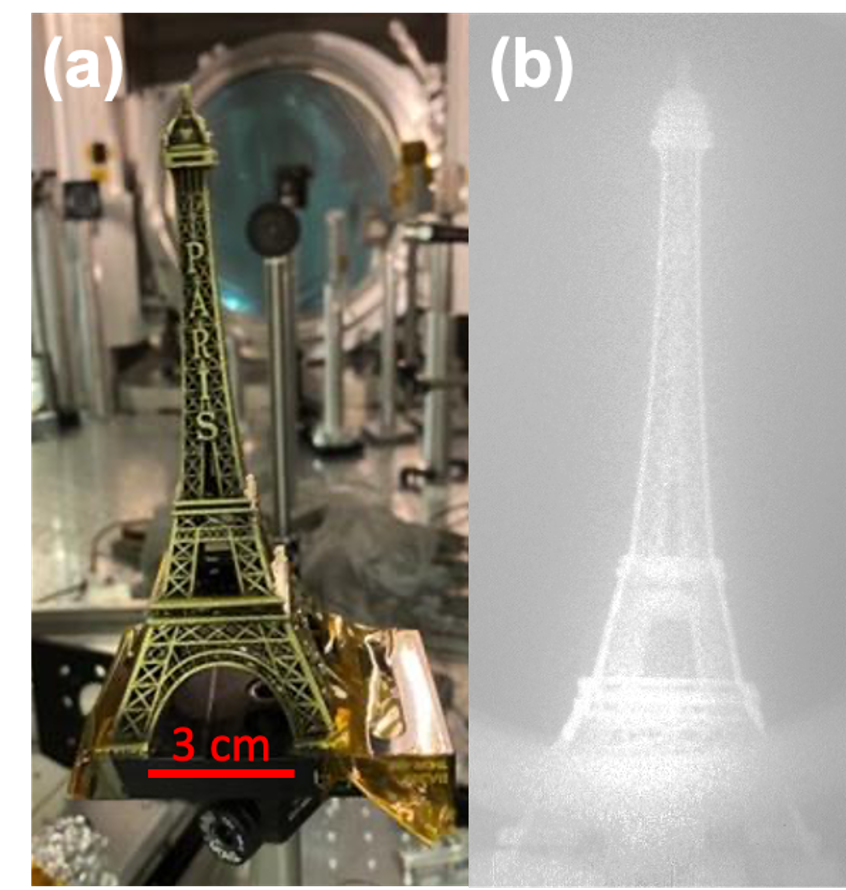}
    \caption{X-ray point-projection imaging. (a) Photo and (b) X-ray radiography of an Eiffel Tower model positioned in the target chamber. The X-rays are produced from a 6 $\mu$m-thick Al target.}
    \label{fig:X_tower}
\end{figure}

These significant X-ray emissions allowed us to perform radiography.
Figure~\ref{fig:X_tower} (b) shows a point-projection image, taken by a 40 cm $\times$ 40 cm, Varex 1621HE a-Si flat panel detector using a DRZ-high scintillator with 0.5 mm Cu intensifier of an object shown in Figure~\ref{fig:X_tower} (a). The point-projection arrangement is such that the object is magnified by a factor 2 on the detector. The resolution of the detector is $\sim$125 $\mu$m. From the image we can tell that the source size has to be $\ll$ 125 $\mu$m. This is consistent with our 2D PIC simulation results, 
in which we diagnose that the transverse spatial distribution of the electron number density (and thus of the radiated X-rays), temporally integrated over 200 fs, has a FWHM of about 20 $\mu$m.

\section{Simultaneous operation of F1 with F2}
\label{F2}

\begin{figure}[ht]
    \centering
    \includegraphics[width=0.48\textwidth]{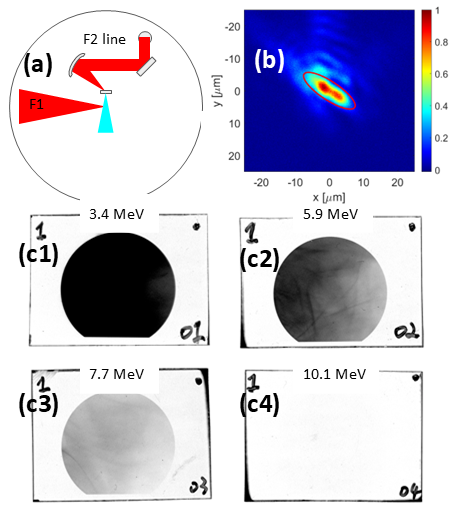}
    \caption{The simultaneous operation of the F1 and F2 beamlines. (a) The setup of using the F2 beamline as the source for proton radiography of plasmas produced by the F1 beamline. (b) The focal spot image of the F2 beamline. The red ellipse has an area of 80.1 $\mu m^2$ and contains 40\% of the total energy. (c1-c4) The raw image of a RCF stack collecting protons produced from the F2 beamline interacting with a 3 $\mu$m Al target.  
    }
    \label{fig:F2-F1}
\end{figure}

We have also demonstrated the simultaneous operation of the F1 and F2 beamlines. As shown in Fig.~\ref{fig:F2-F1}~(a), in this configuration, the F2 beam was focused 8 cm away from the TCC, where the F1 beam was focused. The F2 beam irradiated a thin Al target (i.e., 3 $\mu$m), enabling proton radiography \cite{schaeffer2023proton} of the plasmas generated with the F1 beamline. The average on-target energy of the F2 beamline was about 6 J, with a focal spot shown in Fig.~\ref{fig:F2-F1}~(b). The red ellipse contains 40\% of the energy and has an area of 80.1 $\mu m^2$. Considering a measured pulse duration for F2 of 45 fs, we can estimate the intensity of the F2 beamline during this simultaneous operation to be of the order of $6.7\times10^{19}$ W/cm$^2$. It should be noted, that
both the F2 focal spot and pulse duration were not optimal, due to the fact that both the Dazzler (optimizing the spectrum and phase of the laser pulse) and the deformable mirror (optimizing the wavefront) were geared for optimum F1 operation. Indeed, currently the Apollon facility has one deformable mirror in the beamline. Future plans to employ an additional deformable mirror would enable the optimization of both focal spots independently. A sample energetic proton beam generated by F2, as collected by a RCF stack, is shown in Fig.~\ref{fig:F2-F1} (c1-c4). 
Although the energy is not high, the beam spatial distribution is quite smooth, showing that it is suitable for proton radiography.
The cutoff energy of the proton beam was between 7.7 and 10.1 MeV.
Note that the timing of F2 relative to F1 can be adjusted by an optical delay line from -1 to 5 ns.
Note also that since F1 and F2 are splits from the same beamline, there is no intrinsic jitter between the two. 

\section{Prospect}
\label{prospects}

\begin{figure}[ht]
    \centering
    \includegraphics[width=0.49\textwidth]{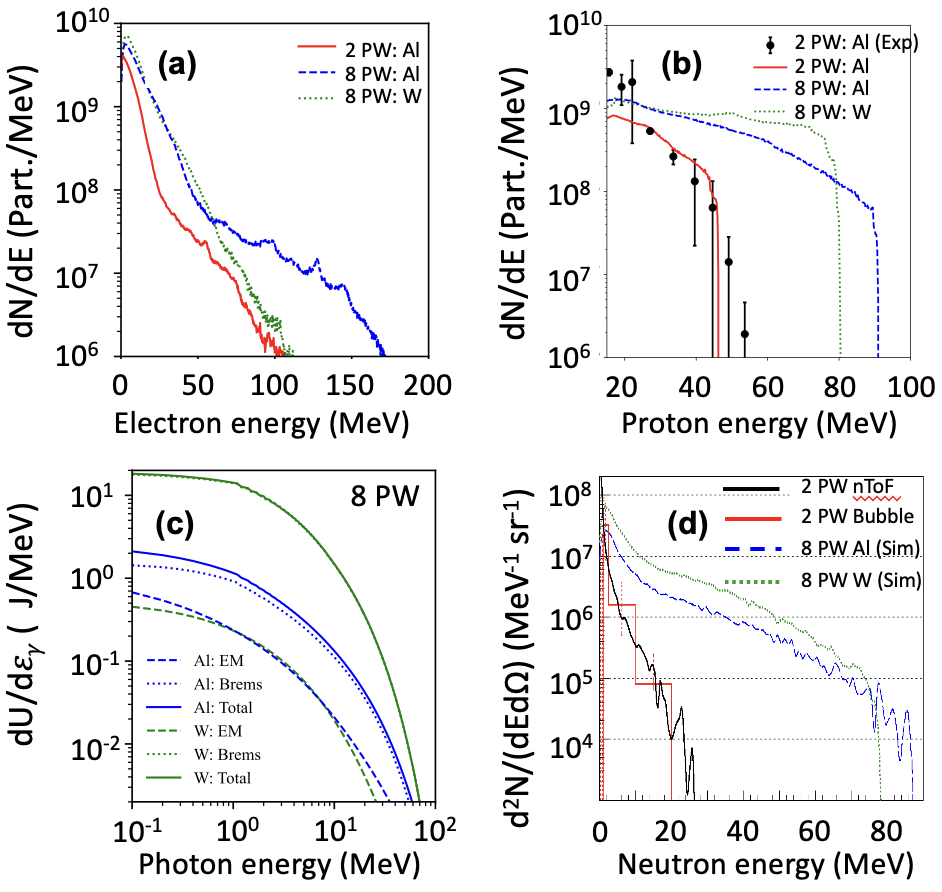}
    \caption{Energy spectra of (a) electrons, (b) protons, (c) radiated photons (both induced by EM-fields and also from Bremsstrahlung) and (d) neutrons, when we will boost the laser power to 8 PW in 2025. In (a), (b), and (d), the 2 PW case is also presented for comparison; while in (c), no 2 PW case is presented for readability. For the 8 PW cases, both Al and W targets are used. Note that these simulations are conducted with a fixed ionization of the pre-plasma and the target (being Z=3 for Al, and Z=14 for W), since dynamical ionization simulations are very demanding, but will be conducted prior to the upgrade.}
    \label{fig:spec_pros}
\end{figure}

Since the Apollon F1 beam is planned to reach the 8 PW level in 2025, we have performed simulations at the 8 PW level for two target materials (a light one, Al, and a high-Z one, W). The results of these simulations are shown in Fig.~\ref{fig:spec_pros}. 
From panel (a), we clearly see that more hot electrons will be produced. As a result, in panel (b), the proton cutoff energy is expected to be almost doubled (from about 50 MeV to about 90 MeV). These enhancements in electrons and protons will then favor both the X-ray generation and neutron production. Panel (c) shows the effect of the target material on X-ray generation, i.e., with the same laser power (of 8 PW). By changing the target material from Al to W, we see that for the W-case, the Bremsstrahlung radiation is enhanced by almost one order-of-magnitude, while the EM-induced radiation stays about the same compared to the Al-case. In panel (d), the neutron cutoff energy, using the Al target with 8 PW laser power, can be boosted to about 90 MeV, which is about 3 times higher than the neutron energy recorded during this commissioning. These radiations should allow the community of users to field a wide range of experiments.

\section{Conclusion}
\label{conclusion}

The Apollon laser facility is a users’ facility open to researchers worldwide. The present paper details how its main beamline (i.e., F1) deployed in the SFA area underwent successful commissioning at a 2 PW level. A number of diagnostics were employed to characterize the laser-target coupling, as well as derive the produced X-rays, protons and neutrons.

We demonstrated that TNSA proton beams with cutoff energies around 50 MeV could be produced from 6 to 8 $\mu$m thick Al foils. 
The emissions of ions, X-rays and neutrons that were recorded show an overall performance that is highly consistent with what has been reported by similar international facilities \cite{danson2019petawatt,doria2020overview,hong2021commissioning}, despite the existence of a pre-pulse. Upcoming experimental campaigns will be carried out to improve the laser contrast in order to allow interactions with ultra-thin targets \cite{yao2024enhanced}. 
We have also demonstrated the possibility of simultaneous operation of F1 with F2 beamline.

The next phase will be the upgrade of the pump energy of the last amplifier to the 700 J level and optimization of the pulse bandwidth/duration of Apollon, which is scheduled for 2025, in order to increase the peak power capacity of the system to the 8 PW level.

\bigskip
\textbf{Acknowledgement}
The authors acknowledge the national research infrastructure Apollon and the LULI staff for their technical assistance. This work was supported by funding from the European Research Council (ERC) under the European Unions Horizon 2020 research and innovation program (Grant Agreement No. 787539, Project GENESIS), by CNRS through the MITI interdisciplinary programs and by IRSN through its exploratory research program. This work was also supported by the French Agence Nationale de la Recherche, under project ANR LIOR (ANR-24-CE05-6070). The project was also made possible thanks to the credits of the Hubert Curien Maimonides program (No. 50046QF) made available by the French Ministry of Europe and Foreign Affairs, the French Ministry of Higher Education and Research and the Israel Ministry of Science, Technology and Space, grant \#6111. We acknowledge the financial support of the IdEx University of Bordeaux / Grand Research Program "GPR LIGHT", and of the Pazy Foundation Grant No. 435/2023. The computational resources of this work were supported by the National Sciences and Engineering Research Council of Canada (NSERC) and Compute Canada (Job: pve-323-ac, PA).

\bigskip
\textbf{\textcolor{black}{Author Contributions Statement}}
Conceptualization, J.F.; 
methodology, J.F., F.M., Y.A., F.P., L.L., N.L., L.R., S.P., P.A., Q.D., S.N.C., A.F., S.M., T.V. and A.A.; 
software, W.Y., M.G., E.d'H., F.E.H. G.M., and L.G.; 
investigation, W.Y., R.L., I.C., T.W., J.F., A.Bec., A.Bel., D.C., M.C., E.C., M.D., C.E., Y.H., H.L., L.L., E.V., D.C.G., I.P., D.P., D.Ma., D.Mi., F.P., N.L., Y.A., Ar.B., Au.B., S.M., Fr.G. P.R., and Fa.G.; 
data curation, H.L., Y.H., W.Y., I.C., J.F., E.F., R.L., F.P, L.L., L.R., P.R., F.T., C.B. and D.C.G.; 
writing---original draft preparation, W.Y., J.F., R.L., E.F., I.C., F.P., L.L., L.R. and L.G.; 
writing---review and editing, all. 
All authors have read and agreed to the published version of the manuscript.

\bigskip
\textbf{{Competing Interests Statement}}

The authors declare no competing interests.

\bigskip
\textbf{{Data availability}}

All data needed to evaluate the conclusions in the paper are present in the paper.
Experimental data and simulations are archived on servers at LULI laboratory and are available from the corresponding author upon reasonable request.

\bigskip
\textbf{Code availability}
The code used to generate the data shown in Fig.~\ref{fig:spec_rcf}, ~\ref{fig:spec_elec}, ~\ref{fig:RCF_angular} (c), ~\ref{fig:Dose_Map_RPL} (b), ~\ref{fig:spec_x}, and ~\ref{fig:spec_pros} (a-c) is \textsc{SMILEI}. The code used to generate the data shown in Fig.~\ref{fig:Neutron_Comparison_Spectra}, ~\ref{fig:Neutron_Simulated_Angular_Distribution}, and ~\ref{fig:spec_pros} (d) is \textsc{Geant4}. Both codes are detailed in the texts.

\bibliography{refs}
\bibliographystyle{apsrev4-2-titles}

\end{document}